\documentclass[astrosymb,trackchanges,twocolumn]{aastex7}
 \hypersetup{linkcolor=red,citecolor=blue,filecolor=cyan,urlcolor=magenta}

\usepackage{amsmath}
\shorttitle{X-ray flaring and quiescent states of K-type stars}
\shortauthors{G. Singh et al.}
\received{August 20, 2025}
\revised{.........}
\accepted{..........}
\submitjournal{AAS}
\usepackage{subfigure}

\begin{document}

\title{Quiescent and flaring states of three active stars: V834 Tau,  LQ Hya, and BY Dra }

\author[orcid=0009-0002-6580-3931,sname='Singh']{Gurpreet Singh}
\affiliation{Aryabhatta Research Institute of Observational Sciences (ARIES), Manora Peak, Nainital 263002, India}
\email[show]{gurpreet@aries.res.in}  

\author[orcid=0000-0002-4331-1867, sname='Pandey']{Jeewan C. Pandey} 
\affiliation{Aryabhatta Research Institute of Observational Sciences (ARIES), Manora Peak, Nainital 263002, India}
\email[show]{jeewan@aries.res.in}

\author[orcid=0000-0001-8620-4511, sname='Karmakar']{Subhajeet Karmakar} 
\affiliation{NASA Goddard Space Flight Center, 8800 Greenbelt Rd, Greenbelt, MD 20771, USA}
\email{subhajeet.karmakar@nasa.gov}

\begin{abstract}
We present a detailed X-ray study of the quiescent and flaring coronae of three active main-sequence K-type stars, V834 Tau, LQ Hya, and BY Dra, using \textit{XMM–Newton} observations. The quiescent coronae are well described by two-temperature thermal plasma models, with cool and hot components at 0.26–0.30 keV and 0.93–1.01 keV, respectively. Despite similar coronal temperatures, X-ray luminosities (10$^{29.18\mbox{--}29.75}$ erg s$^{-1}$) and overall abundances, the relative emission measures of the cool and hot components differ among the stars. High-resolution spectroscopy reveals significant iron depletion by factors of 5-10 relative to photospheric values, and an inverse first-ionization-potential effect in all three stellar coronae. Six energetic flares are detected, with peak temperatures of 30 -- 133 MK and released energies of $0.6\mbox{--}4.2\times$10$^ {33}$ erg, classifying them as superflares. Most flares exhibit decay times roughly twice their rise times, although one event shows a decay phase nearly twenty times longer than its rise. Time-resolved spectroscopy and loop scaling laws yield flare parameters consistent with previous studies of active stars. LQ Hya displays recurrent superflares at the same rotational phase across observations separated by six months, suggesting a long-lived, complex magnetic field structure. These results provide insights into the magnetic activity and flare energetics in active stars, and their implications for stellar and exoplanetary environments.

\end{abstract}


\section{Introduction}
The coupling of rotation and convection in late-type stars drives diverse magnetic activities including flares, starspots, coronal mass ejections, and activity cycles \citep[][]{,2004A&ARv..12...71G,2009A&ARv..17..251S}.
 The studies have shown that the X-ray activity increases with faster rotation and saturates at a certain point. The rotation period at which this saturation occurs is a function of the star's spectral class (and thus its mass). For a $1\text{ M}_{\odot}$ star, saturation is typically observed at rotation periods shorter than 4 days \citep{2003A&A...397..147P, 2011ApJ...743...48W}.
While all stars within the saturation regime exhibit identical X-ray to bolometric luminosity ratio ($L_X/L_{\text{bol}} \approx 10^{-3}$), their underlying coronal properties, such as elemental abundances, emission measures, and filling factors, are different,  making these stars more interesting to study. 
Coronal elemental abundances often show patterns linked to first ionization potential (FIP). In many active stars, an apparent inverse-FIP effect has been reported \citep[][]{2015LRSP...12....2L}, but in most cases this is based on assuming solar photospheric abundances. When the star’s actual photospheric composition is used, the effect often weakens or disappears \citep[][]{2004A&A...416..281S,2009A&A...505..299S,2015A&A...577A..93P}, highlighting the need to study it carefully to understand whether coronal heating and fractionation in active stars truly differ from the Sun.

 The quiescent state of active stars is often interrupted by intense flaring episodes, which heat the coronal plasma to extreme levels.
Flares arise from magnetic reconnection, in which the reconnection process accelerates charged particles and heats the plasma to several million kelvin within a coronal loop, producing intense emission across the full electromagnetic spectrum, from radio wavelengths to X-rays \citep{2017ApJ...851...91N}. The frequency and intensity of these eruptive events are related to the star's rotation and age, as faster-rotating younger stars possess more complex internal dynamos \citep{2014ApJ...792...67C}.  In addition to probing extreme plasma conditions, flares are often accompanied by the ejection of magnetised plasma from the corona, known as a coronal mass ejection (CME). Unlike the solar case, where CMEs can be directly observed through imaging, evidence for stellar CMEs remains indirect. 

Beyond their astrophysical significance, stellar flares and CMEs play a crucial role in exoplanetary science, as their high-energy radiation can strongly affect planetary atmospheres and influence habitability \citep{2019ApJ...881..114Y, 2021MNRAS.500L...1A, 2024A&A...688A.138P}. Consequently, characterising flare statistics is essential for understanding stellar activity and assessing the long-term stability of planetary environments.

\begin{table*}
 \caption{Log of observations from XMM-Newton.}
 \label{tabl:obslog}
\resizebox{\textwidth}{!}{
    \begin{tabular}{cccccccc}
    \hline
    \hline
         Star Name &Observation ID  & Detector & Mode/Filter        & Start time                & Exposure  & Source region & Background region\\ 
                   &                &  &                  & YYYY-MM-DD hh:mm:ss       & ks        &   ( arcsec )        & ( arcsec )  \\
    \hline
         V834 Tau  &  0785141001    &EPIC-PN &Small Window/Thick  & 2016-08-18 23:13:21       & 15.5      & 40            & 39.7, 35.9\\
                   &                & RGS    &                    & 2016-08-18 23:07:23       & 15.9      & &\\
         BY Dra    &  0785140501    &EPIC-PN &Small Window/Thick  & 2016-04-29 10:22:07       & 22.9      & 42            & 40.7, 48.6\\
                   &                & RGS    &                    & 2016-04-29 10:16:09       & 23.3      & &\\
         LQ Hya  (Epoch-I)  &  0148880101    & EPIC-PN& Small Window/Medium & 2003-05-09 04:57:04       & 63.4      & 50            & 40.5, 40.6\\
                            &                & RGS    &                     & 2003-05-09 04:51:43       & 71.2      & &\\
      LQ Hya  (Epoch-II) &  0148880301    &EPIC-PN &Small Window/Medium & 2003-11-22 08:30:10       & 15.5      & 50            & 30, 30\\
                         &                & RGS    &                    & 2003-11-22 08:24:05       & 15.9      & &\\
    \hline
    \end{tabular}
    }\\
\centering{\scriptsize Background region is the composite of two circular regions near the source with minimal source contamination. }
\end{table*}

In this paper, we investigate the quiescent and flaring activity of three K-type active dwarfs in the saturation regime, namely V834 Tau, BY Dra, and LQ Hya. These stars are located within 20 pc of the Sun; thus, they are bright in X-rays, making them suitable for detailed time-resolved spectral analysis to understand the evolution of coronal properties.

V834 Tau is a K3V-type star \citep[][]{2009A&A...508.1313F}, with rotation period of 3.936$\pm$0.003 d \citep[][]{1995AJ....110.2926H}. It is a young star with an age of 49$\pm$37 Myr \cite[][]{2009ApJ...698.1068P}. Although optical observations have been conducted, the star lacks a detailed individual study, and its X-ray properties have not yet been investigated.

BY Dra is the prototype of its class (BY Draconis variables).  It is a spectroscopic binary of K4Ve+K7Ve type with orbital period of \textbf{$\sim$5.98 d} and eccentricity of 0.3 \citep[][]{2012mnras.419.1285h}. The rotation period of the active stellar component \textbf{(i.e. primary)} is determined to be 3.83~d  \citep[][]{1992A&AS...96..497P}. The first X-ray flare detection from BY Dra occurred in 1986 using EXOSAT data; however, detailed analysis was limited by poor X-ray data statistics \citep[][]{1986A&A...156...95D}.  The age of BY Dra is estimated to be  $\sim$1-2 Gyr \citep[][]{2012mnras.419.1285h}, making it younger than the Sun.  Despite being a prototype of its own kind of variables, no detailed study on its corona has been carried out thus far.

LQ Hya is a K2V type star \citep[][]{1986AJ.....92.1150F} with rotation period of \textbf{$\sim$ 1.6 d}
\citep[][]{1993A&A...276..345J}. It is also a very young star with the age of 51.9$\pm$17.5 Myr \citep[][]{2011mnras.410..190T}. LQ Hya is previously reported to exhibit X-ray flaring activity based on ROSAT and ASCA data \citep{2001A&A...371..973C}. 

The paper is organized as follows: Section \ref{sec:obs} summarizes X-ray observations and data reduction procedures. Section~\ref{sec:analysisandresults} details the temporal and spectral analysis, presenting results for quiescent and flaring states and including abundance analysis from high-resolution X-ray data. Section~\ref{sec:RGS} describes the RGS spectral analysis, while Section~\ref{sec:loop} presents the flare loop modeling. Section \ref{sec:discuss} discusses the results, and finally, Section~\ref{sec:conc} provides the conclusions.

\begin{figure*}
\centering
\subfigure[V834 Tau Obs ID: 0785141001 ]{\includegraphics[scale=0.23]{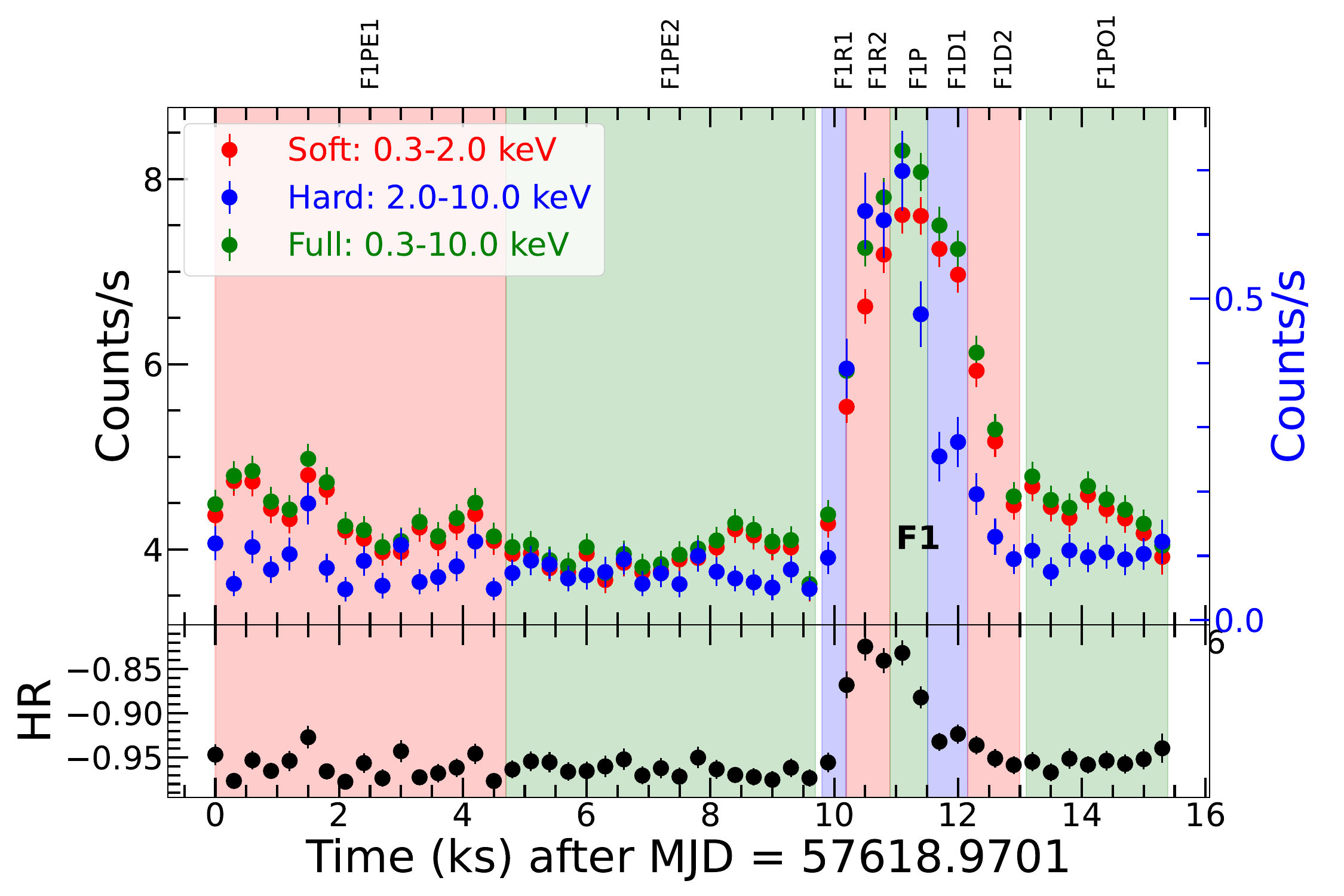}}
\subfigure[BY Dra Obs ID: 0785140501 ]{\includegraphics[scale=0.23]{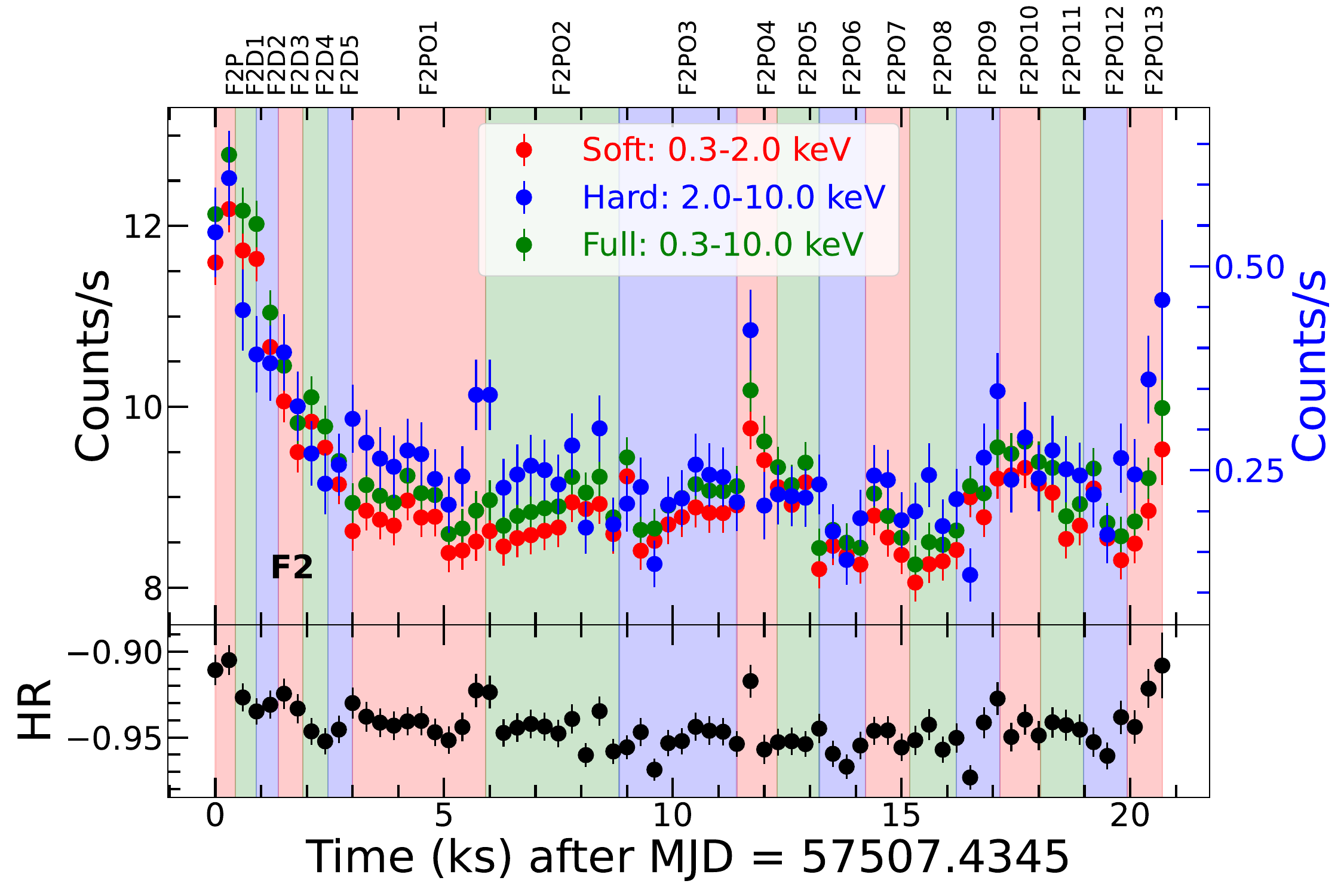}}\\
\subfigure[LQ Hya Obs ID: 0148880101 ]{\includegraphics[scale=0.23]{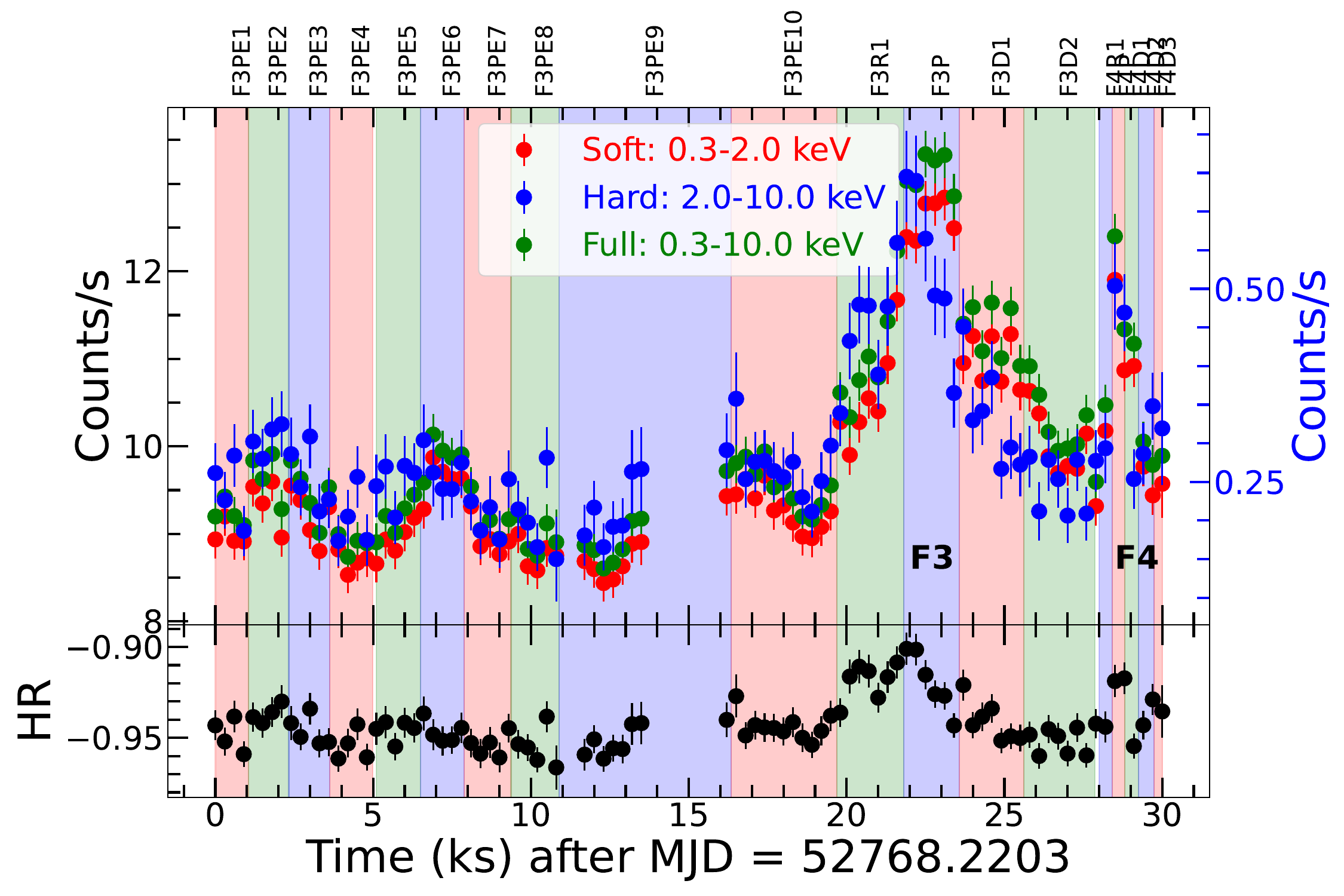}}
\subfigure[LQ Hya Obs ID: 0148880301 ]{\includegraphics[scale=0.23]{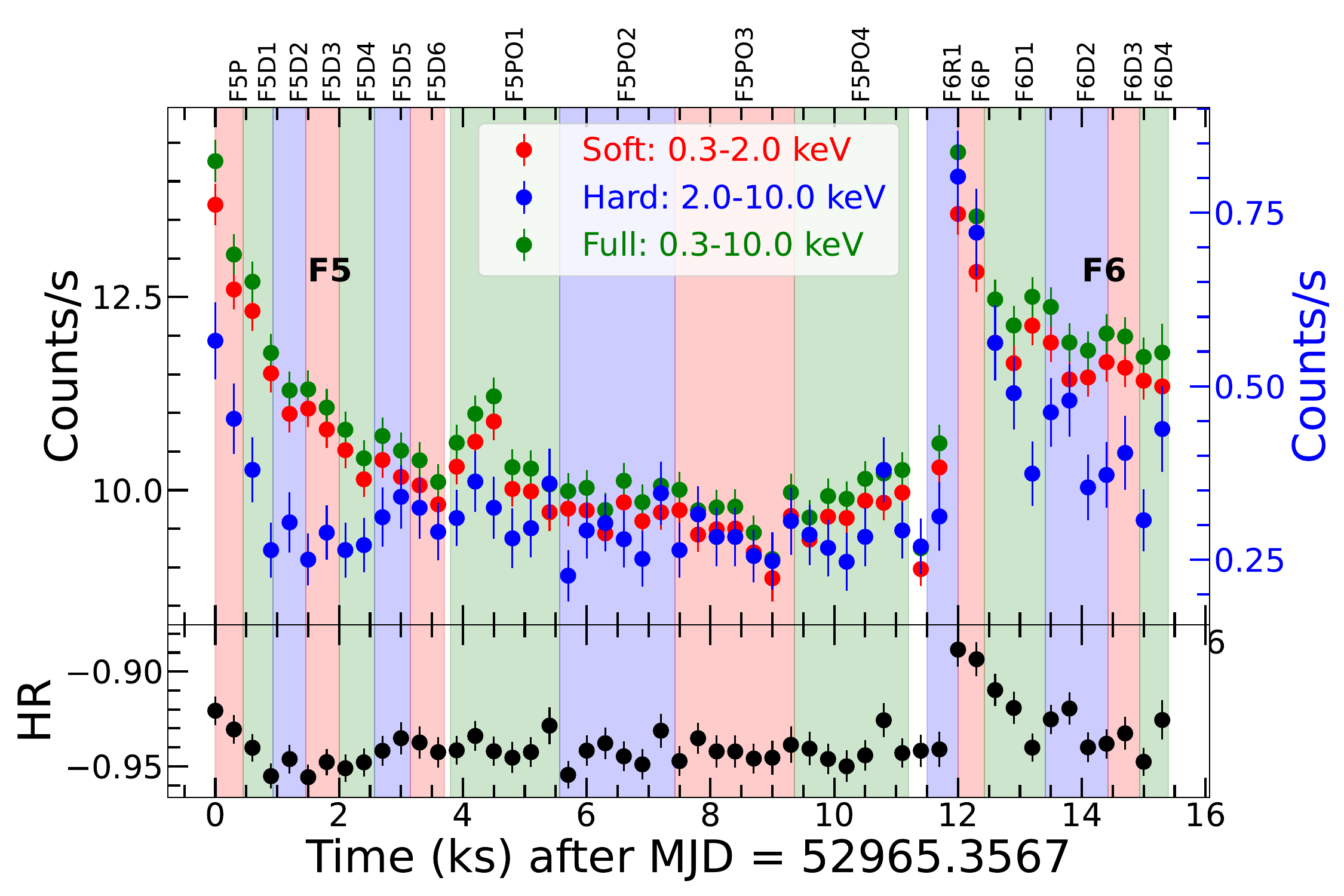}}
\caption{X-ray light curves for sample stars in S (red), H(blue), and F(green) bands, along with the hardness ratio in the bottom panels. The left Y-axis represent the count rate for the S and F bands, whereas the right Y-axis represents the H band count rate for each top panel of each Figure. The segments for time-resolved spectra extraction are shown as shaded regions, with corresponding labels given at the top of each plot.}
    \label{fig:sampleLCS}
\end{figure*}

\section{Observations and Data reduction}\label{sec:obs}
All three stars were observed by the XMM-Newton observatory in 2003 and 2016. A  log of observations is given in Table \ref{tabl:obslog}. The data were retrieved from the XMM-Newton science archive and reduced using SAS v 21.0.0. Detailed data reduction steps followed standard procedures outlined in SAS threads\footnote{\href{https://www.cosmos.esa.int/web/xmm-newton/sas-threads}{https://www.cosmos.esa.int/web/xmm-newton/sas-threads}}. In summary, the raw EPIC-PN data were processed and calibrated using the {\sc epproc} command. The presence of soft proton flares was checked by generating a background light curve in the energy range of 10.0--12.0 keV. Periods of high background were excluded by generating a Good Time Interval file (GTI) using the {\sc tabgtigen} task in SAS. After applying GTI to calibrated data, source and background light curves and spectra were generated using the {\sc evselect} task. These were extracted from on-source counts within circular regions centered on the source and from two nearby source-free background regions. The selected source and background regions are also given in Table \ref{tabl:obslog}. The resulting spectra were tested for pile-up effects using {\sc epatplot} task. We did not find any significant pile-up in these observations. The response matrices were generated using {\sc rmfgen} and {\sc arfgen} commands. The source, background, and detector response matrices were grouped together using {\sc grppha} tool within {\sc HEASOFT}. X-ray light curves were also corrected for detector response, area normalization, dead pixels, and background subtraction using the {\sc epiclccorr} task in {\sc SAS}.  In order to remove fast stochastic variability (less than 5 min) and improve the statistics, the X-ray light curves were binned to a 300 s time bin size using the {\sc lcurve} tool in {\sc HEASOFT}. X-ray spectra were fitted using {\sc xspec} within {\sc HEASOFT}.

The RGS data were reduced using {\sc rgsproc}, which performs all the basic reduction procedures and generates calibrated event files, spectra, light curves, and source-extraction masks. For background-flare checks, we have generated light curves from CCD chip 9 of RGS1 and RGS2 using {\sc evselect}. The GTI files were generated where necessary, using {\sc tabgtigen} command. These new GTI files were applied to re-generate the spectra from RGS1 and RGS2 using {\sc rgsproc}. The resulting background flare-free spectra from RGS1 and RGS2 were then grouped together to ensure each energy bin contained a minimum of 20 counts. 

\section{Analysis and Results}\label{sec:analysisandresults}
\subsection{X-ray light curves}

X-ray light curves were generated in three energy bands: a soft band (S: 0.3--2.0 keV), a hard band (H: 2.0--10.0 keV), and a full band (F: 0.3--10.0 keV). 
To investigate temperature variations in X-ray emission, we calculated the Hardness Ratio (HR) = $\frac{H-S}{H+S}$. Variations in the hardness ratio (HR) enable distinguishing between quiescent and flaring states, as the spectrum becomes harder during flaring episodes and the HR curve follows the flare light curve. Figure~\ref{fig:sampleLCS} displays the X-ray light curves for the H, S, and F bands in the upper three panels, with the corresponding HR curves shown in the bottom panels. The epoch-I observations of LQ Hya were significantly affected by soft proton background flares after the $\sim$30 ks of observation start, and a small proton flare episode was noted between 14 - 16 ks. Therefore, further analysis was performed by excluding these background flaring episodes. 

We identified a total of six flares in current observations of three stars, four of which occurred on LQ Hya. The flares were marked as F1...6 in  Figure \ref{fig:sampleLCS}. The e-folding rise ($\tau_r$) and decay  ($\tau_d$) times for the flares were calculated using the exponential rise and exponential decay model, which was shifted to match the quiescent value during preflare and postflare times. The derived rise and decay times are given in Table \ref{tab:loopparam}. The decay times of these flares were found to be approximately twice as large as the rise times, demonstrating a typical fast rise and slow decay in all the flares except for F6, which shows a very steep rise that was $\sim$16 times shorter than the decay time. 

\subsection{X-ray Spectral Analysis}
The quiescent (i.e., pre-/post-flaring) and flaring segments of the light curve were analyzed separately for the X-ray spectroscopy. Figure \ref{fig:sampleLCS} shows the different time segments for the quiescent and flaring states. The rise, peak, decay, pre-flare, and post-flare phases were marked by FiRj, FiP, FiDj, FiPEj, and FiPOj, respectively, where i,j = 1,2...... The segments were chosen in such a way that each segment has a similar area under the curve in the full band X-ray light curve. This ensures each segment follows the same statistics for the fit.
 Each spectrum was fitted with multiple thermal plasma models {\sc apec} \citep{2001ApJ...556L..91S}. The hydrogen column density ($N_H$) was found to be not well constrained by the model, so it was fixed to the value $10^{18}$ atom/$cm^2$ for further analysis. We adopted the solar photospheric abundance (Z$_\odot$) from \cite{1989GeCoA..53..197A}. For all three stars, the quiescent state spectra were well fitted by a two-temperature plasma model with consistent temperatures (T) near  $\sim$0.3 and $\sim$1.0 keV,  whereas the emission measures (EM) in the quiescent states were found to be slightly different but well within 2.5 $\sigma$ level for different segments of the quiescent state. To obtain the flaring component of the plasma, the spectra of flaring segments were fitted with a three-temperature plasma model, keeping the temperatures and emission measures of the first two {\sc apec} components fixed at their quiescent values, which is equivalent to the flare emission with the quiescent level removed. This approach lets us determine a single "effective" temperature and emission measure for the flaring plasma.
  
\begin{figure*}
    \centering
    \subfigure[V834 Tau]{\includegraphics[width=0.98\columnwidth, trim=1cm 1.6cm 2cm 4cm, clip=true, angle=0]{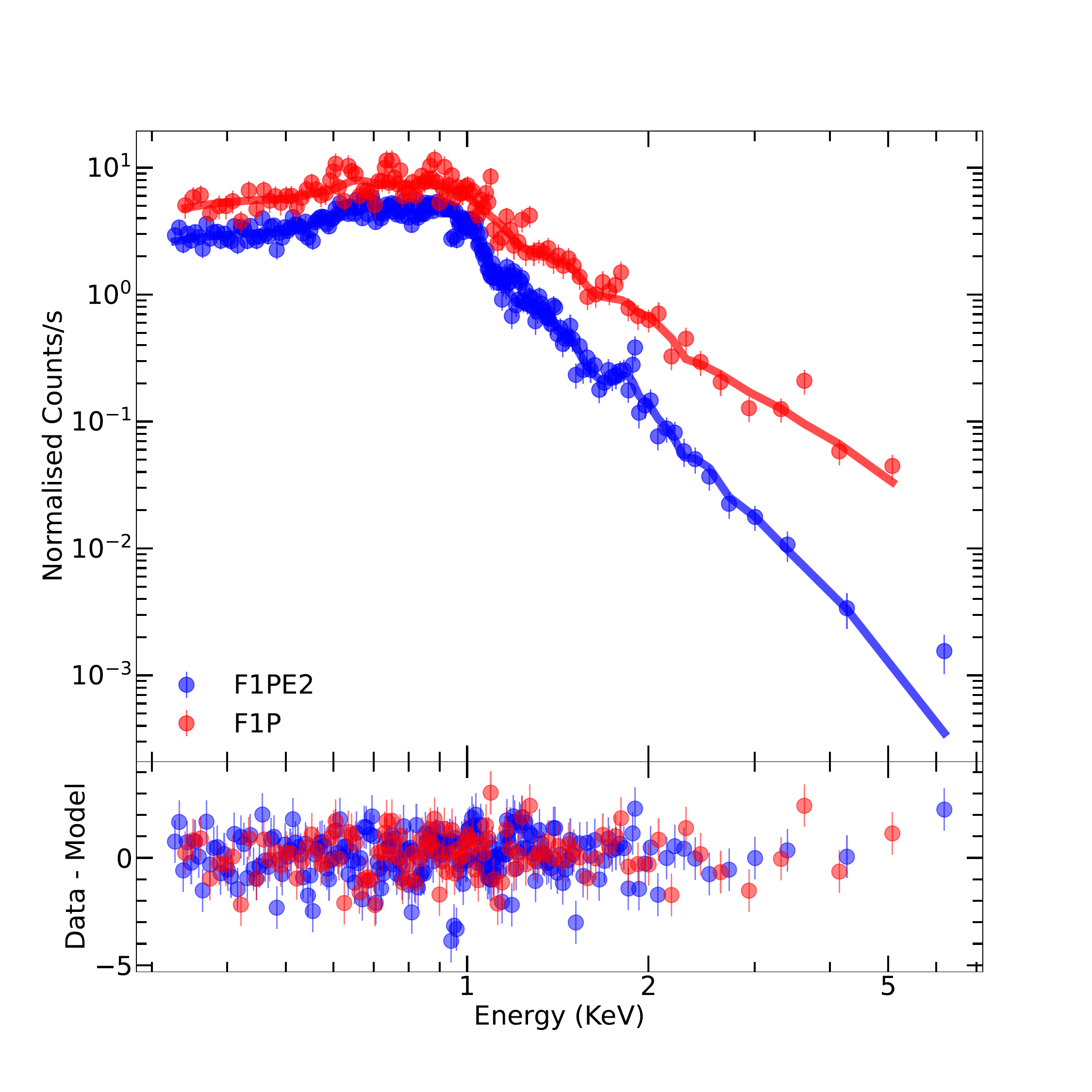}}
        \subfigure[BY Dra]{\includegraphics[width=0.98\columnwidth, trim=1cm 1.6cm 2cm 4cm, clip=true, angle=0 ]{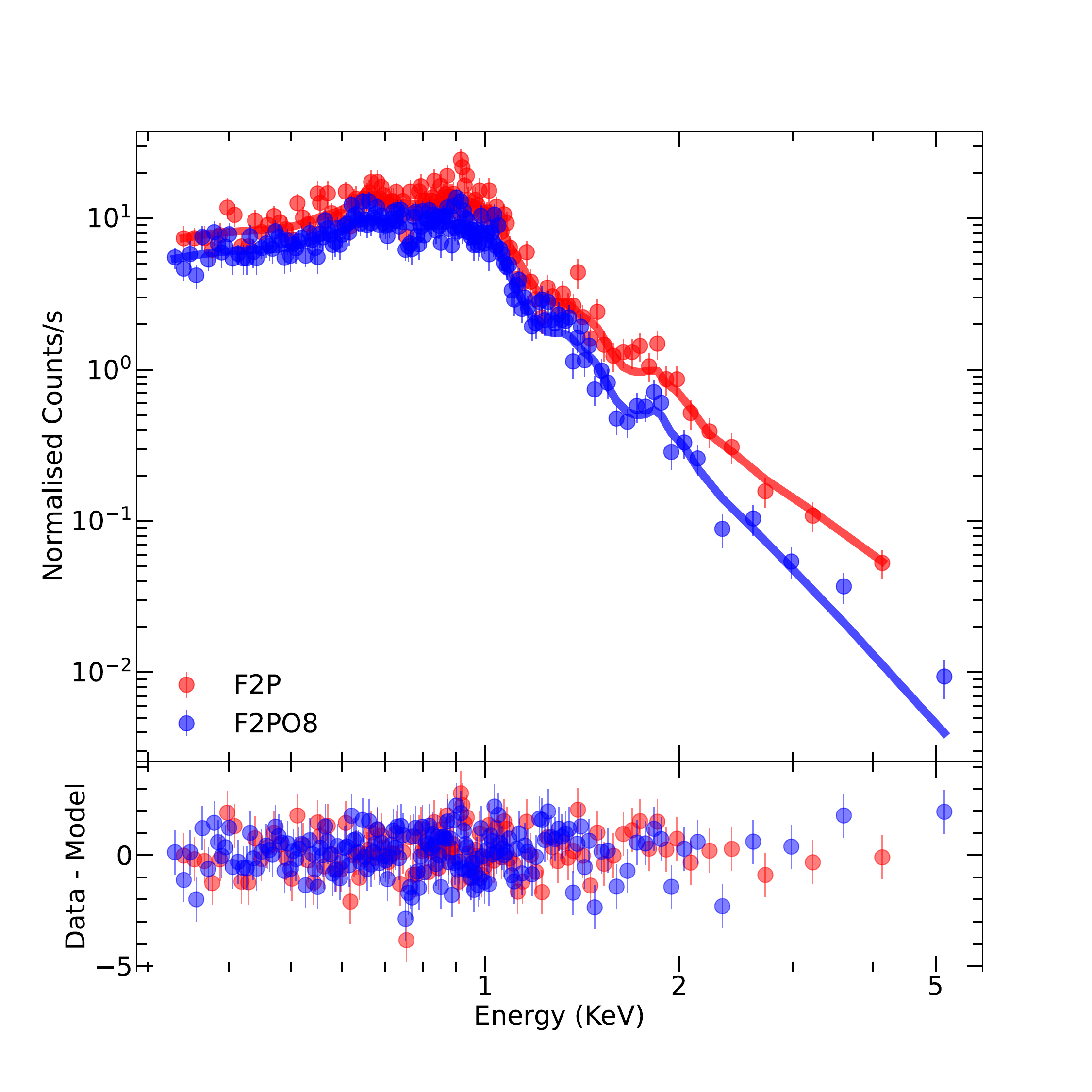}}
    \subfigure[LQ Hya (Epoch-I) ]{\includegraphics[width=0.98\columnwidth, trim=1cm 1.6cm 2cm 4cm, clip=true, angle=0]{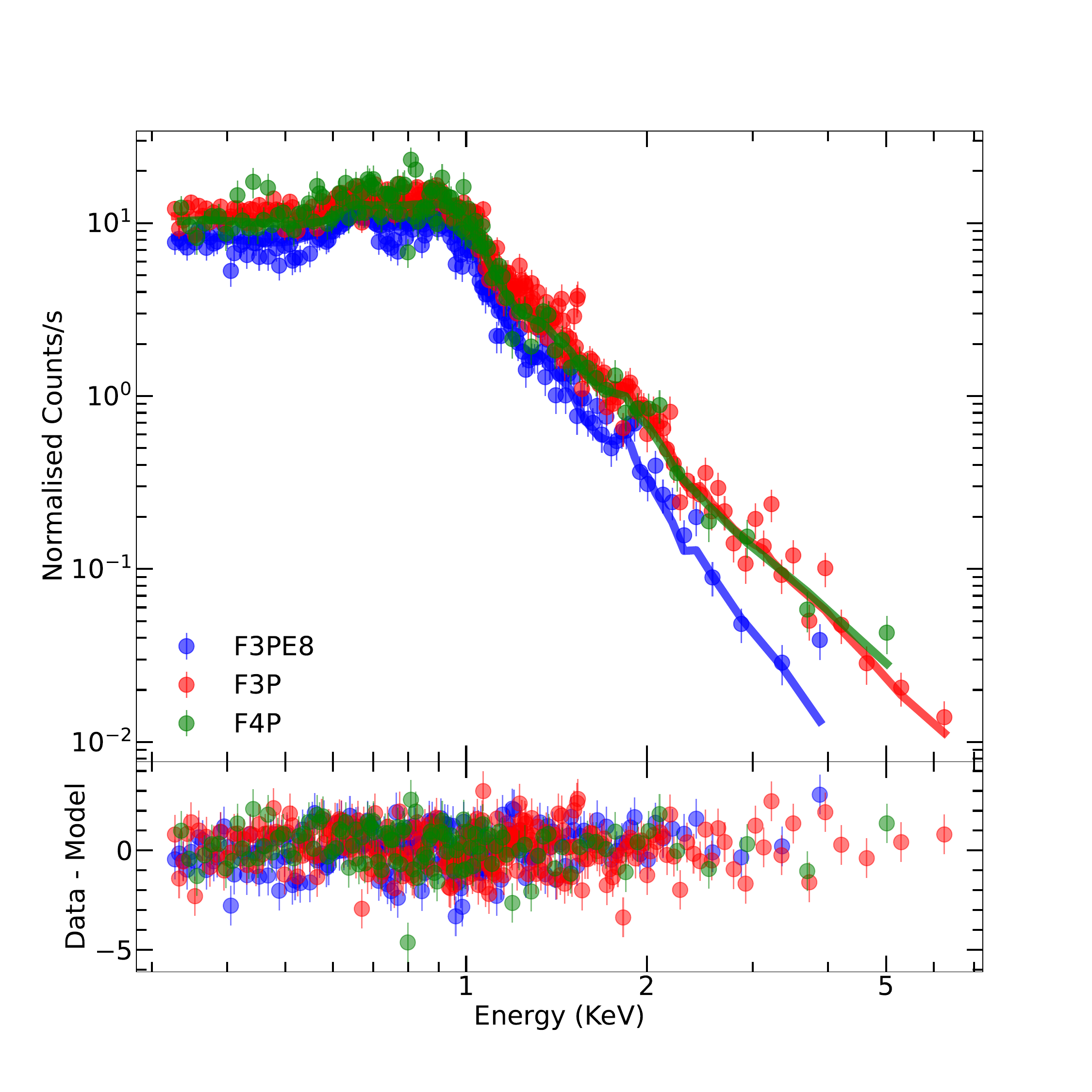}}
    \subfigure[LQ Hya (Epoch-II) ]{\includegraphics[width=0.98\columnwidth, trim=1cm 1.6cm 2cm 4cm, clip=true, angle=0]{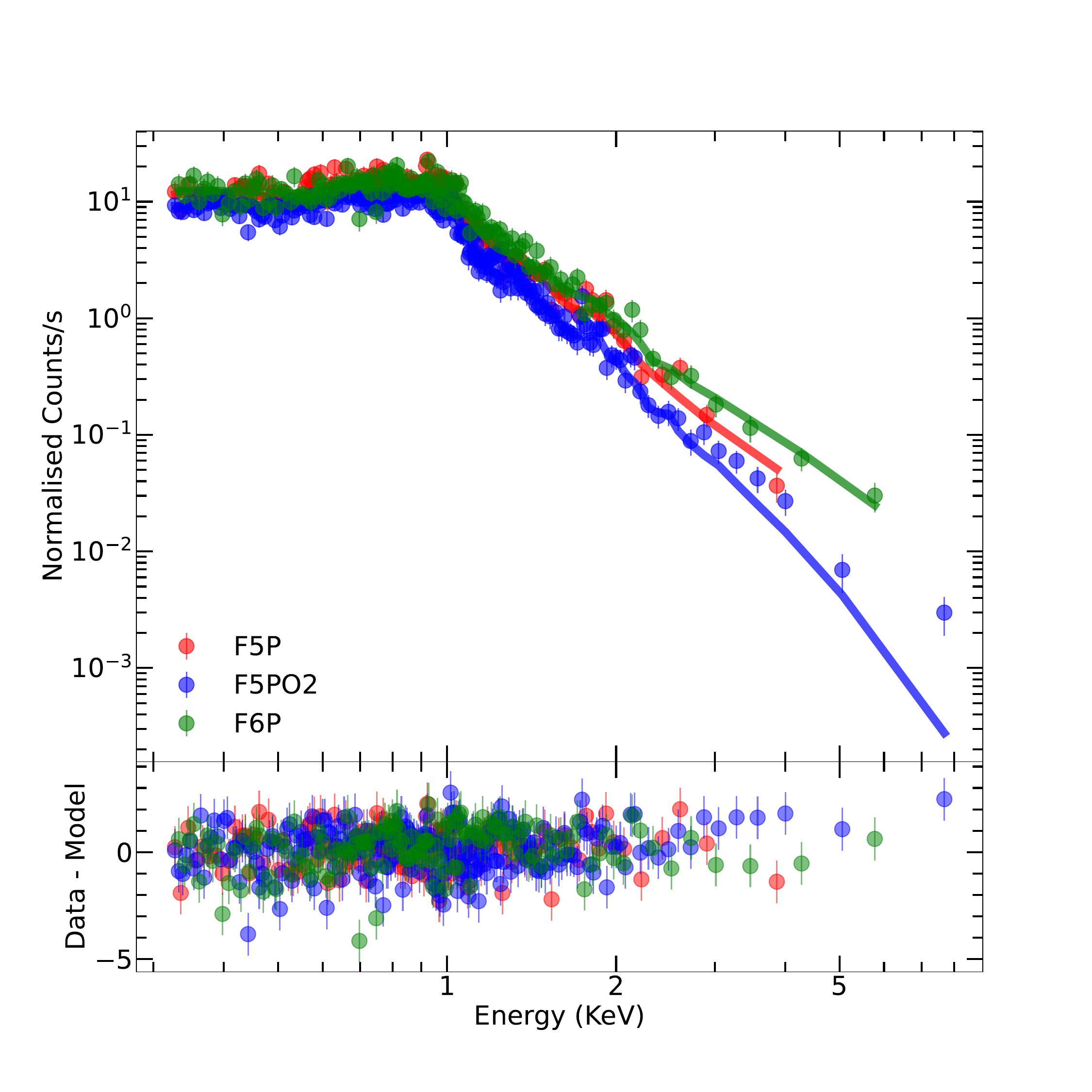}}
    
    \caption{ The X-ray spectra as obtained from the EPIC-PN instrument of three stars, along with the best fit thermal plasma model. The blue colored spectra are for the quiescent phase.  The red/green colors represent spectra during the flare peak. For (a) and (c), only one of many quiescent segments is shown for clarity, while the actual quiescent is the average of all the quiescent segments.}
    \label{fig:bestfit}
\end{figure*}

 The X-ray spectra along the best-fit thermal plasma model and the residuals for the peak flare phase and the quiescent state are shown in Figure \ref{fig:bestfit}, whereas Table \ref{tab:TRS} summarizes the derived best-fit parameters. The X-ray spectra during the peak phase show higher emission than that from the quiescent state. The evolution of the spectral parameters like X-ray luminosity (L$_X$), temperature (T), emission measure (EM), and abundances (Z) is shown in Figure \ref{fig:TRS} for all three stars.

\subsubsection{V834 Tau}
For  V834 Tau, the average temperatures of cool and hot plasma components during the quiescent state were measured to be 0.267$\pm$0.006 and 0.94$\pm$0.01 keV, with the corresponding emission measures of 1.17$\pm$0.06$\times$10$^{52}$ and 9.7$\pm$0.6$\times$10$^{51}$ cm$^{-3}$, respectively. The average quiescent abundances were found to be 0.195$\pm$0.007 Z$_\odot$. The evolution of spectral parameters during the flare is shown in Figure \ref{fig:TRS_v834tau}. The temperature peaked during the rise phase of the flare with a value of 2.6$\pm$0.3 keV, whereas the emission measure peaked simultaneously with X-ray luminosity at a value of 1.6$\pm$0.1$\times 10^{52} $cm$^{-3}$. The abundances were found to peak during the early decay phase of the flare with the peak value of 0.28$^{+0.01}_{-0.02}$ Z$_\odot$, which is significantly higher than that from the quiescent state value. During flare F1, the peak X-ray luminosity reached  3.54$\pm$0.07$\times$10$^{29}$ erg~s$^{-1}$, which is more than double the quiescent state X-ray luminosity. Like X-ray luminosity, the temperature, emission measure, and abundances exhibit a consistent pattern of rising, peaking, and then declining to their pre-flare lowest values.

\subsubsection{BY Dra} \label{sec:ana_bydra}
The average temperatures of cool and hot components during the quiescent states of BY Dra were found to be 0.267$\pm$0.003 and 1.001$\pm$0.006 keV, respectively; whereas the coronal abundances were found in the range of  0.18--0.29 Z$_\odot$, averaging to a value of 0.232$\pm$0.007 Z$_\odot$. The average values of emission measure during the quiescent state were found to be 3.15$\pm$0.0610$^{52}$ and 3.19$\pm$0.07$\times$10$^{52}$ cm$^{-3}$ for cool and hot components, respectively, whereas the X-ray luminosity was found to be 5.39$\pm$0.02$\times$10$^{29}$ erg~s$^{-1}$.

\begin{figure*}
\centering
\subfigure[V834 Tau Obs ID: 0785141001 ]{\includegraphics[width=0.46\textwidth, trim=1cm 2.5cm 3cm 6cm, clip=true, angle=0]{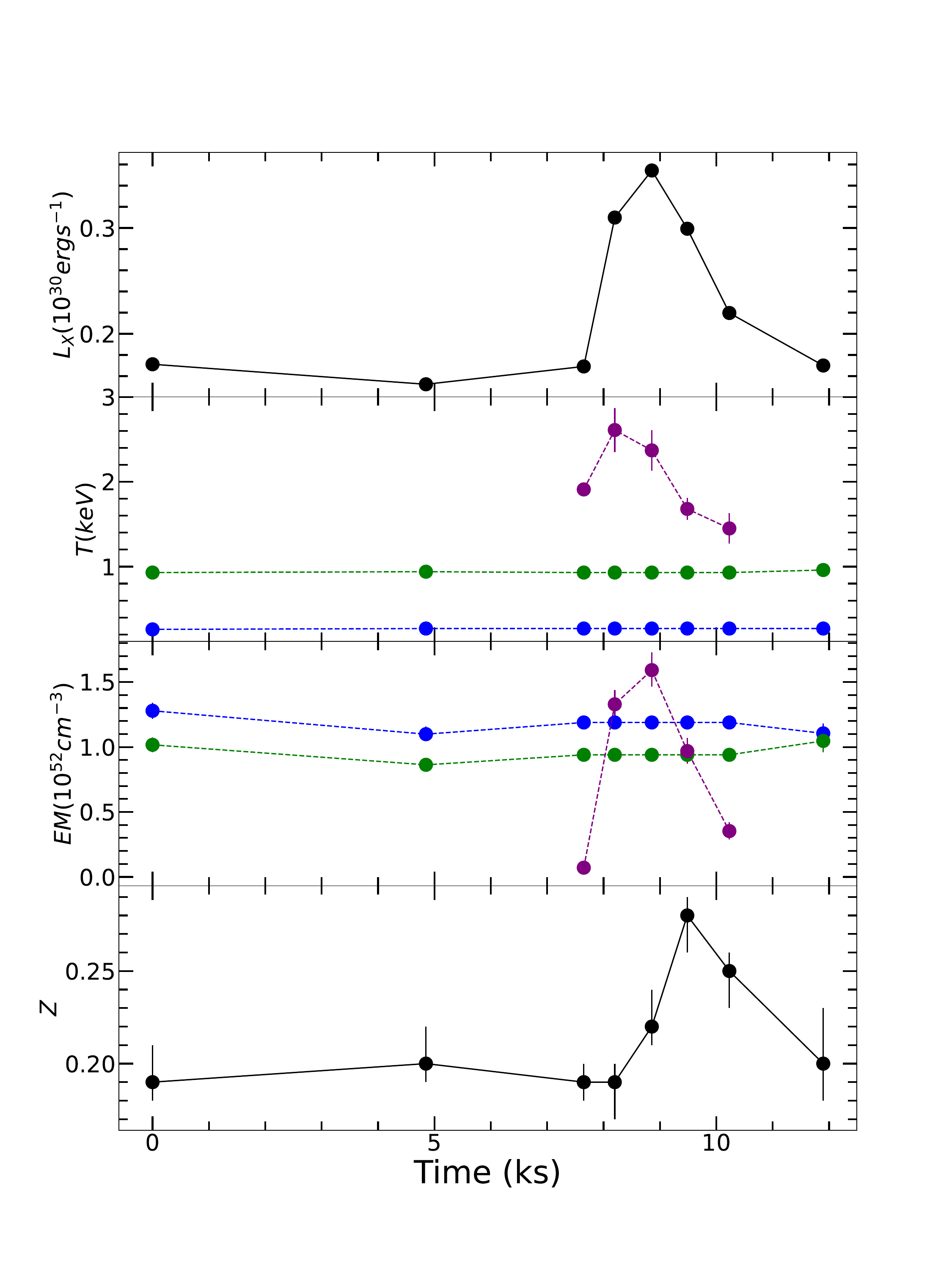} \label{fig:TRS_v834tau}}
\subfigure[BY Dra Obs ID: 0785140501 ]{\includegraphics[width=0.46\textwidth, trim=1cm 2.5cm 3cm 6cm, clip=true, angle=0 ]{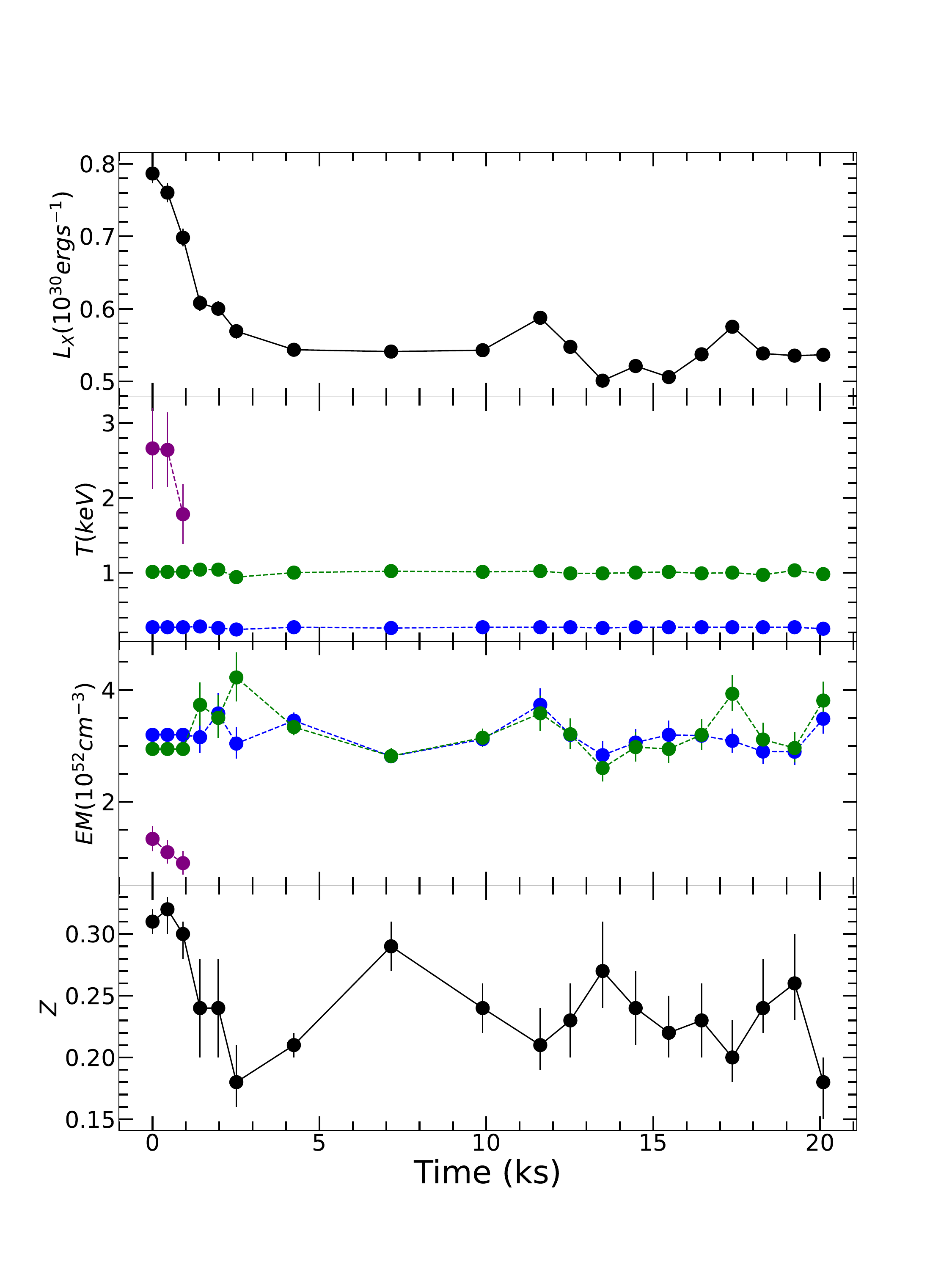} \label{fig:TRS_bydra}}
\subfigure[LQ Hya Obs ID: 0148880101 ]{\includegraphics[width=0.46\textwidth, trim=1cm 2.5cm 3cm 6cm, clip=true, angle=0]{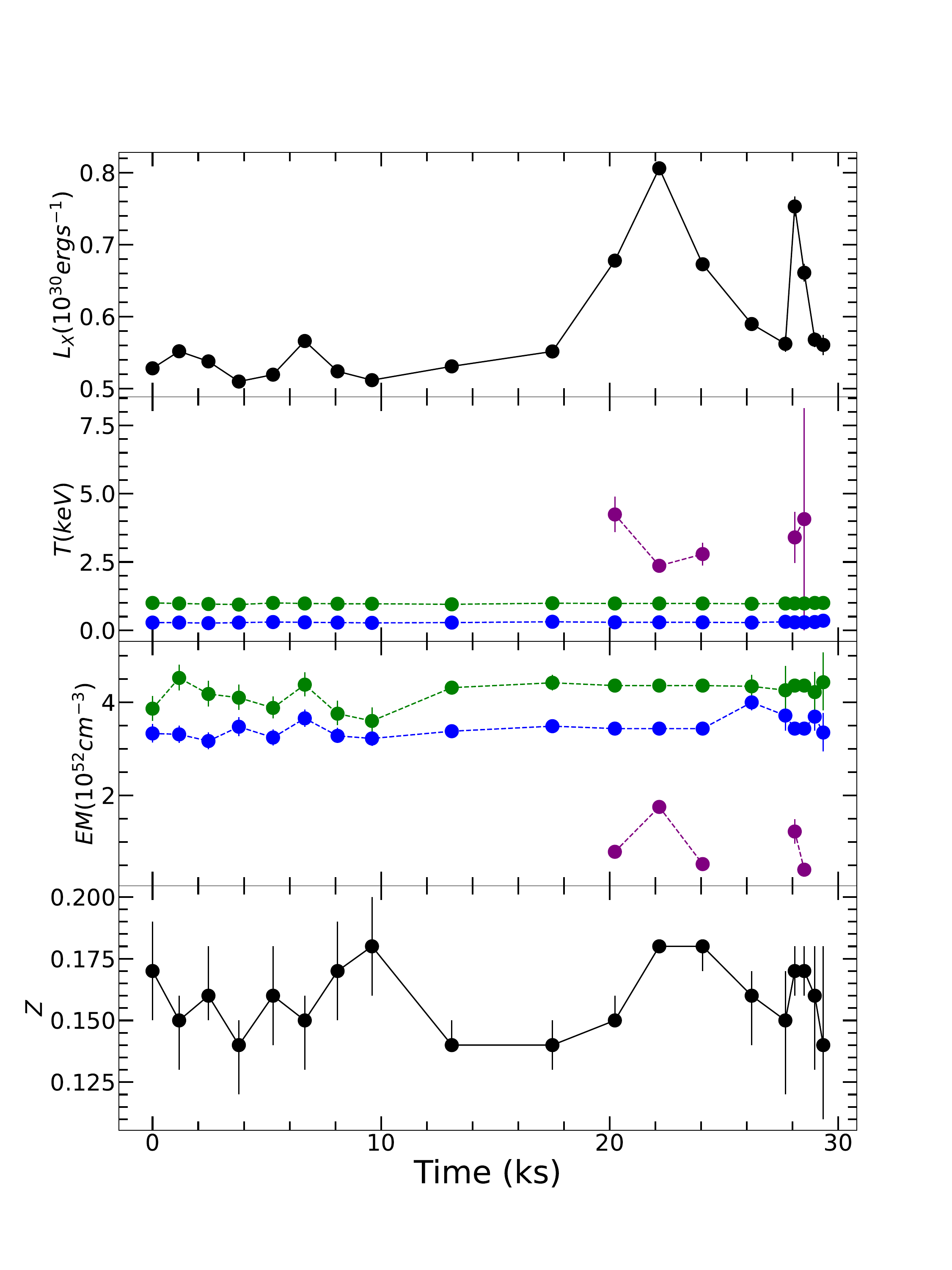} \label{fig:TRS_lqhya1}}
\subfigure[LQ Hya Obs ID: 0148880301 ]{\includegraphics[width=0.46\textwidth, trim=1cm 2.5cm 3cm 6cm, clip=true, angle=0]{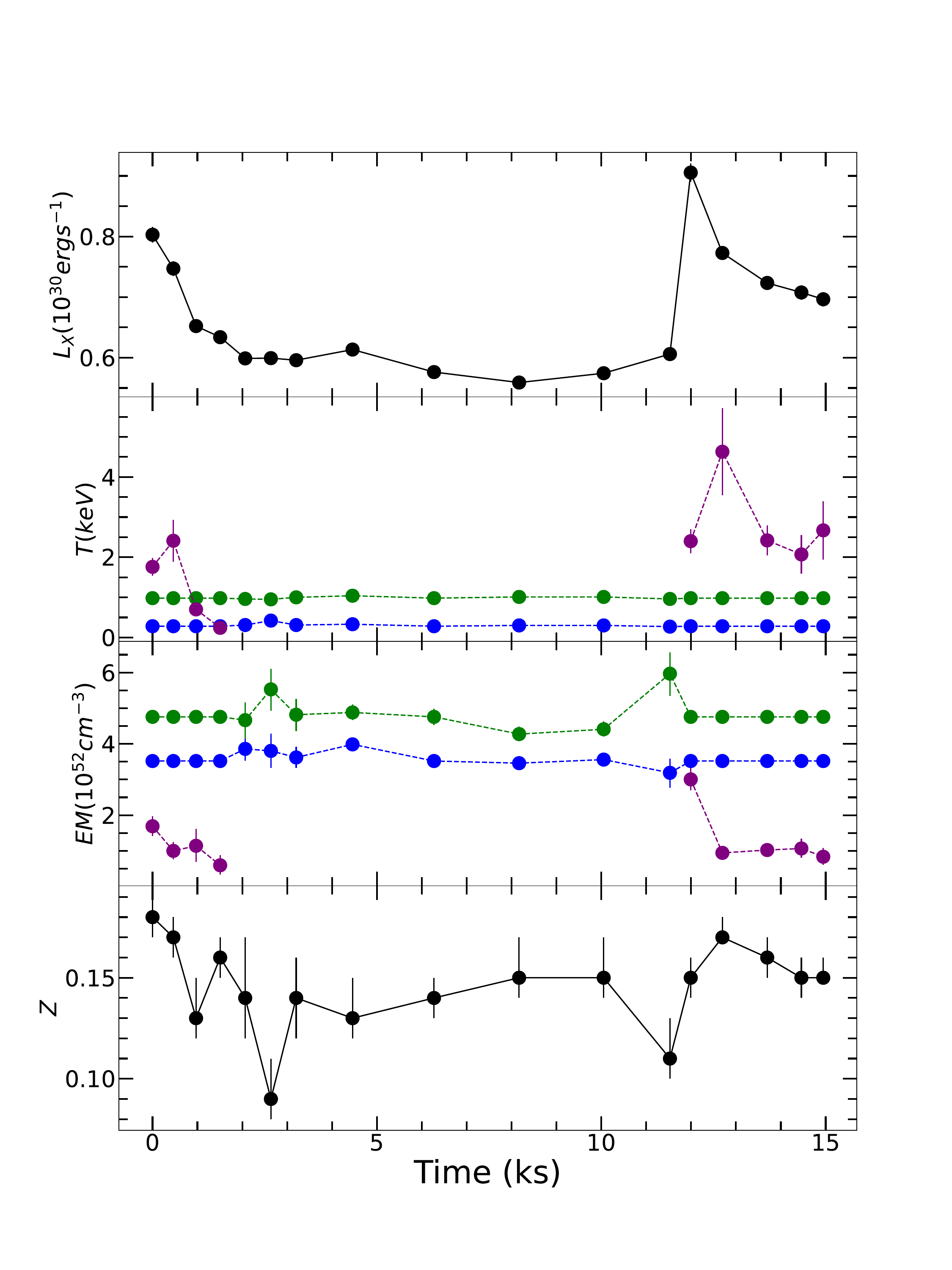}\label{fig:TRS_lqhya2}}
\caption{Evolution of spectral parameters with time. In the first panel of each Figure, the variation of X-ray luminosity in the F band is shown. The second and third panels show the evolution of temperatures and corresponding emission measures. Here, green and blue represent the hot and cool quiescent components, respectively, whereas purple denotes flaring components. In the last panel, the variation of average coronal abundances is shown.}
    \label{fig:TRS}
\end{figure*}

During the flare F2, both temperature and emission measure peaked simultaneously with  X-ray luminosity, with peak flare temperature of 2.66 keV and emission measure of 1.3$\times 10^{52} $cm$^{-3}$. We also note that the coronal abundances during the flare reached a value of $\sim$0.3 Z$_\odot$, whereas the peak X-ray luminosity was found to be more than $\sim$1.5 times that of the quiescent state. The evolution of the spectral parameters is shown in Figure \ref{fig:TRS_bydra}. 
 
\subsubsection{LQ Hya}
 The X-ray light curve of epoch-I of the LQ Hya shows many small-scale fluctuations resembling flare-like events. However, the spectra during these segments were well described by a two-temperature plasma model, with consistent estimates of spectral parameters. Therefore, the entire phase prior to the flare F3 was considered a quiescent state.  We have derived average values of the spectral parameters for epoch-I and epoch-II separately.  In the quiescent state of epoch-I, the average values of cool and hot temperatures were determined to be 0.283$\pm$0.003 keV and 0.974$\pm$0.006 keV, respectively, whereas their corresponding average emission measures were determined to be 3.36$\pm$0.06 $\times$ 10$^{52}$  cm$^{-3}$ and 4.11$\pm$0.08 $\times$ 10$^{52}$ cm$^{-3}$. However, during epoch-II, the average values of temperature for cool and hot components of the quiescent state were found to be 0.303$\pm$0.005 and 1.01$\pm$0.01 with corresponding emission measures of 3.6$\pm$0.1 $\times10^{52}$ cm$^{-3}$ and 4.6$\pm$0.1 $\times10^{52}$ cm$^{-3}$, respectively.  The average coronal abundances were found to be  0.156$\pm$0.005 Z$_\odot$ for epoch-I and   0.143$\pm$0.007 Z$_\odot$ for epoch-II. The average X-ray luminosity during the quiescent state of epoch-I and epoch-II in the 0.3--10.0 keV band was found to be 5.33$\pm$0.02 $\times10^{29}$ and 5.80$\pm$0.03 $\times10^{29}$ erg s$^{-1}$, respectively. All spectral parameters, except for abundances, were found to differ significantly between the quiescent states of epoch-I and epoch-II.
 
A similar approach to the one above was applied to the spectral X-ray data fitting of the flares from LQ Hya.   The evolutions of spectral parameters for flares F3 and  F4, and F5 and F6 are shown in Figures \ref{fig:TRS_lqhya1} and \ref{fig:TRS_lqhya2}, respectively. The peak value of temperature for the flare F3 was found to be 4.24 keV, whereas the peak value of emission measure was found to be 1.8$\times 10^{52}$ cm$^{-3}$. The temperature peaked before the emission measure for flare F3. Towards the end of the observation, a flare F4 was detected and analyzed by dividing the time into four segments. The three-component model was found necessary to fit the spectra of the peak and first-decay segments, in which the first two components were fixed to the quiescent-state values. However, the second (F4D2) and third (F4D3) decay segments were best fit by a two-temperature plasma model with similar temperatures for the cool and hot components in pre-flare states. The peak temperature of the flare F4 was found to be  $\sim$3.4 keV along with a peak emission measure of 1.22$\times 10^{52} $cm$^{-3}$.   The  peak temperatures for the flares F5 and F6 are derived to be  2.41 and 2.40 keV, respectively; whereas the peak emission measures were found to be  1.7$\times$10$^{52}$cm$^{-3}$ and 3.0$\times$10$^{52}$cm$^{-3}$, respectively.
The peak value of coronal abundances was found to be 0.18$\pm$0.01 Z$_\odot$ during flare F5, while for the flare F6, it was found to be similar to those of the quiescent state. 

We also calculated the total X-ray energy ($E_X$) radiated during the flare by integrating the X-ray luminosity in a 0.3--10.0 keV band over the entire flare duration. Flare energies for the four flares from LQ Hya ranged from 0.6 to 4.2 $\times10^{33}$ erg, while the flares from  BY Dra and V834 Tau Tau had energies of 1.03$\times 10^{33}$ and 8.8$\times 10^{32}$ erg, respectively.

\begin{figure*}
    \centering
    \subfigure[V834 Tau]{\includegraphics[width=0.98\columnwidth, trim=1.25cm 1cm 1.5cm 1cm, clip=true, angle=0]{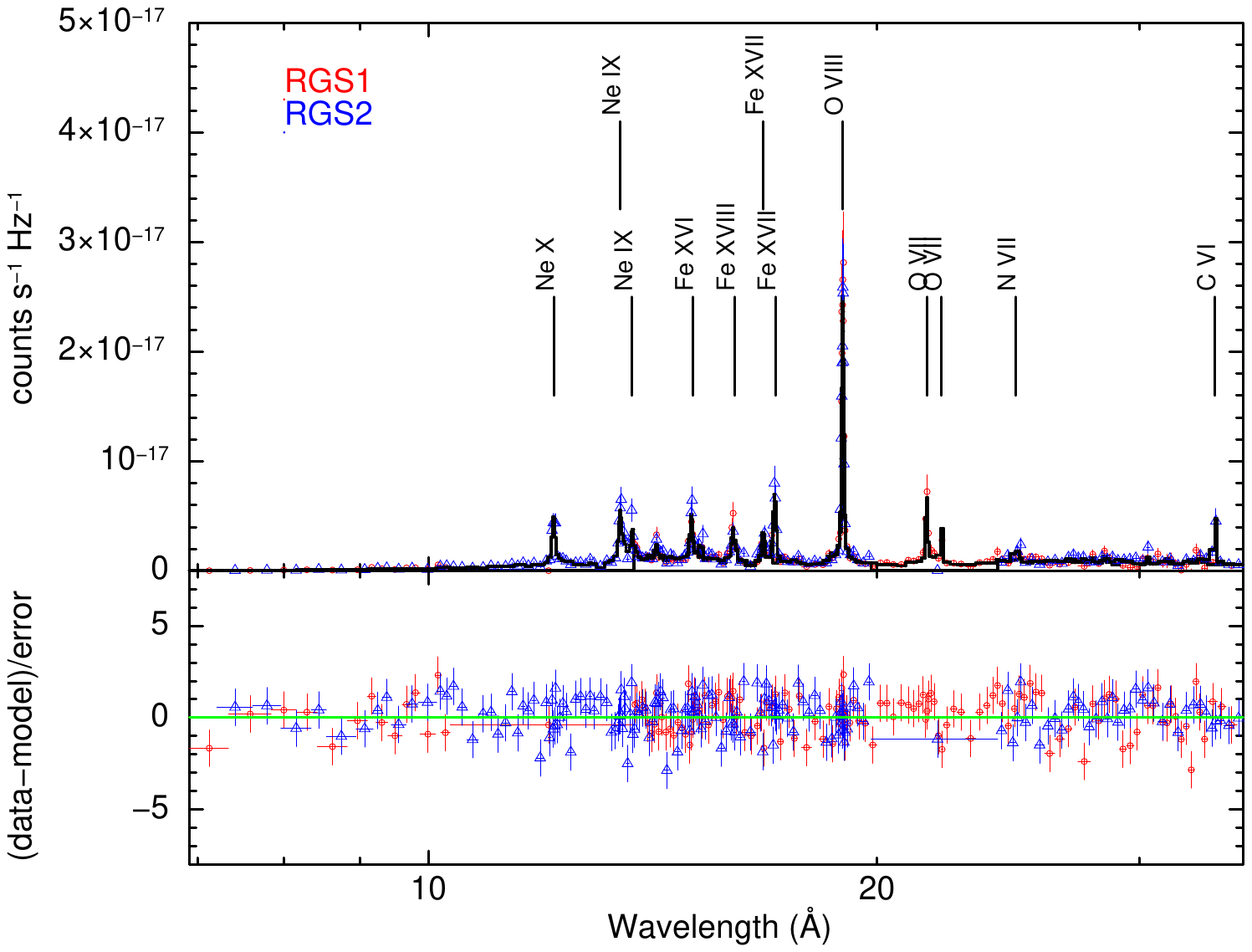}} \label{fig:rgs_v834tau}
        \subfigure[BY Dra]{\includegraphics[width=0.98\columnwidth, trim=1.25cm 1cm 1.5cm 1cm, clip=true, angle=0]{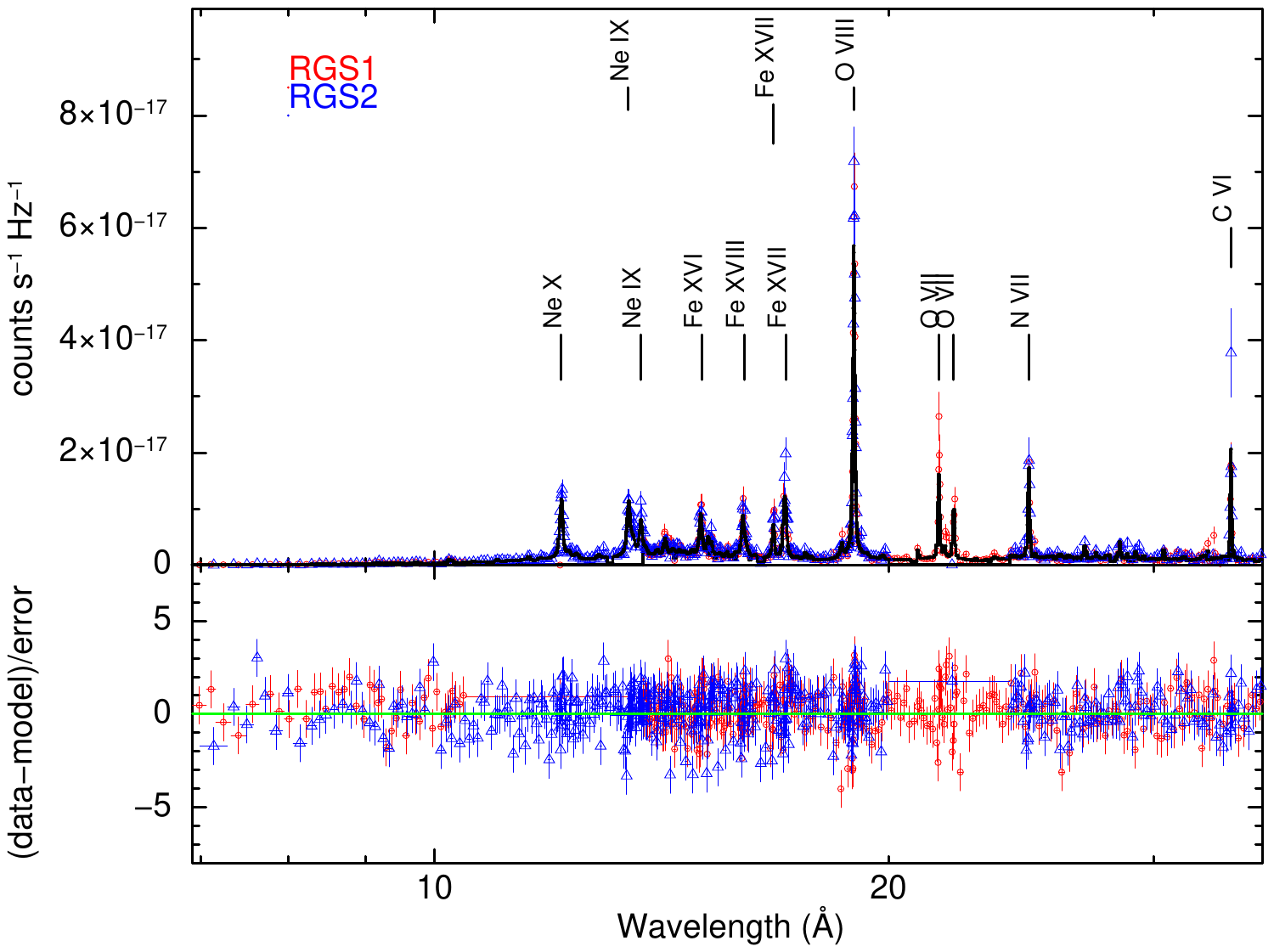}}\label{fig:rgs_bydra}
    \subfigure[LQ Hya (Epoch-I) ]{\includegraphics[width=0.98\columnwidth, trim=1.25cm 1cm 1.5cm 1cm, clip=true, angle=0]{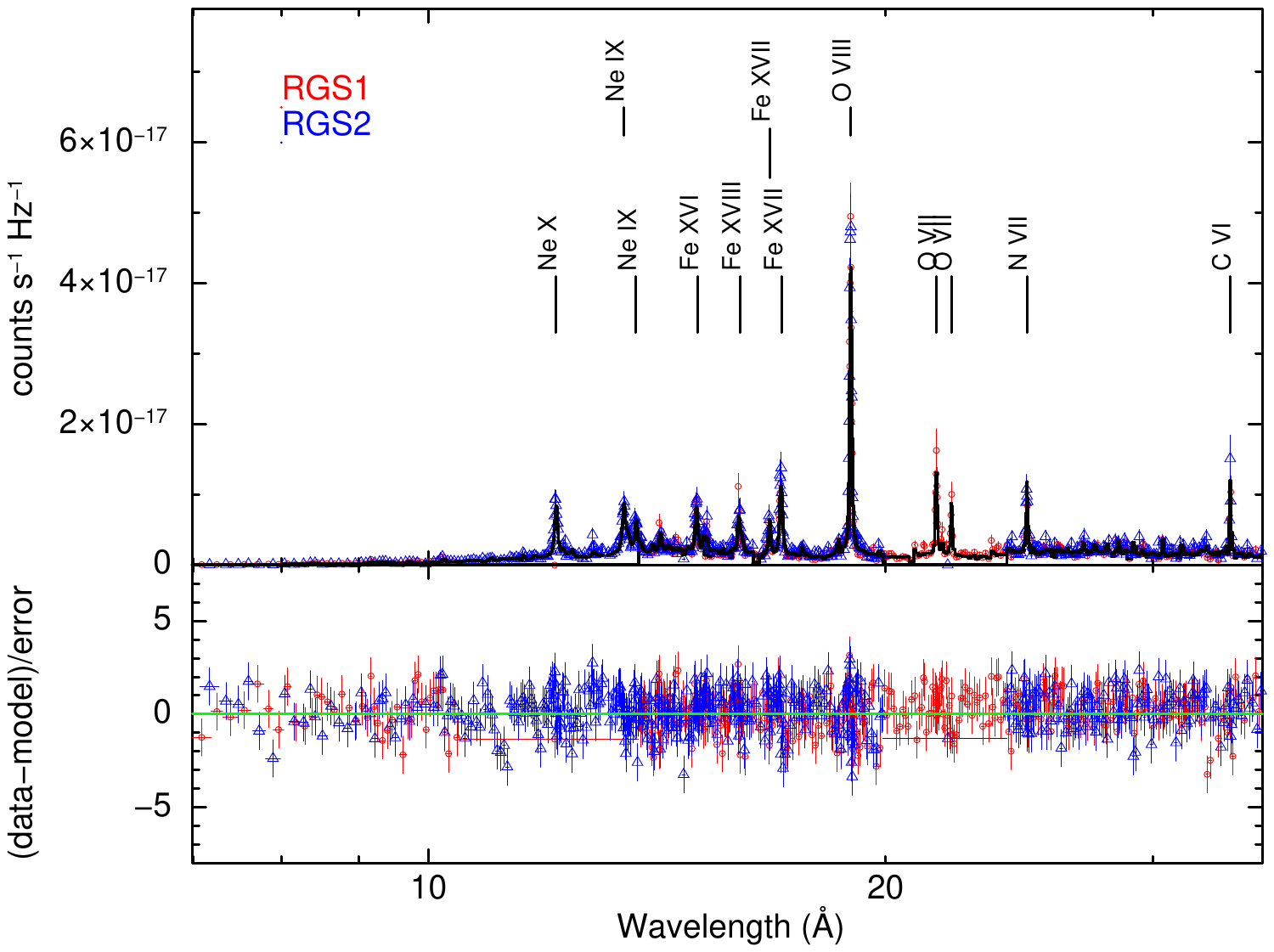}}\label{fig:rgs_lqhya1}
    \subfigure[LQ Hya (Epoch-II) ]{\includegraphics[width=0.98\columnwidth, trim=1.25cm 1cm 1.5cm 1cm, clip=true, angle=0]{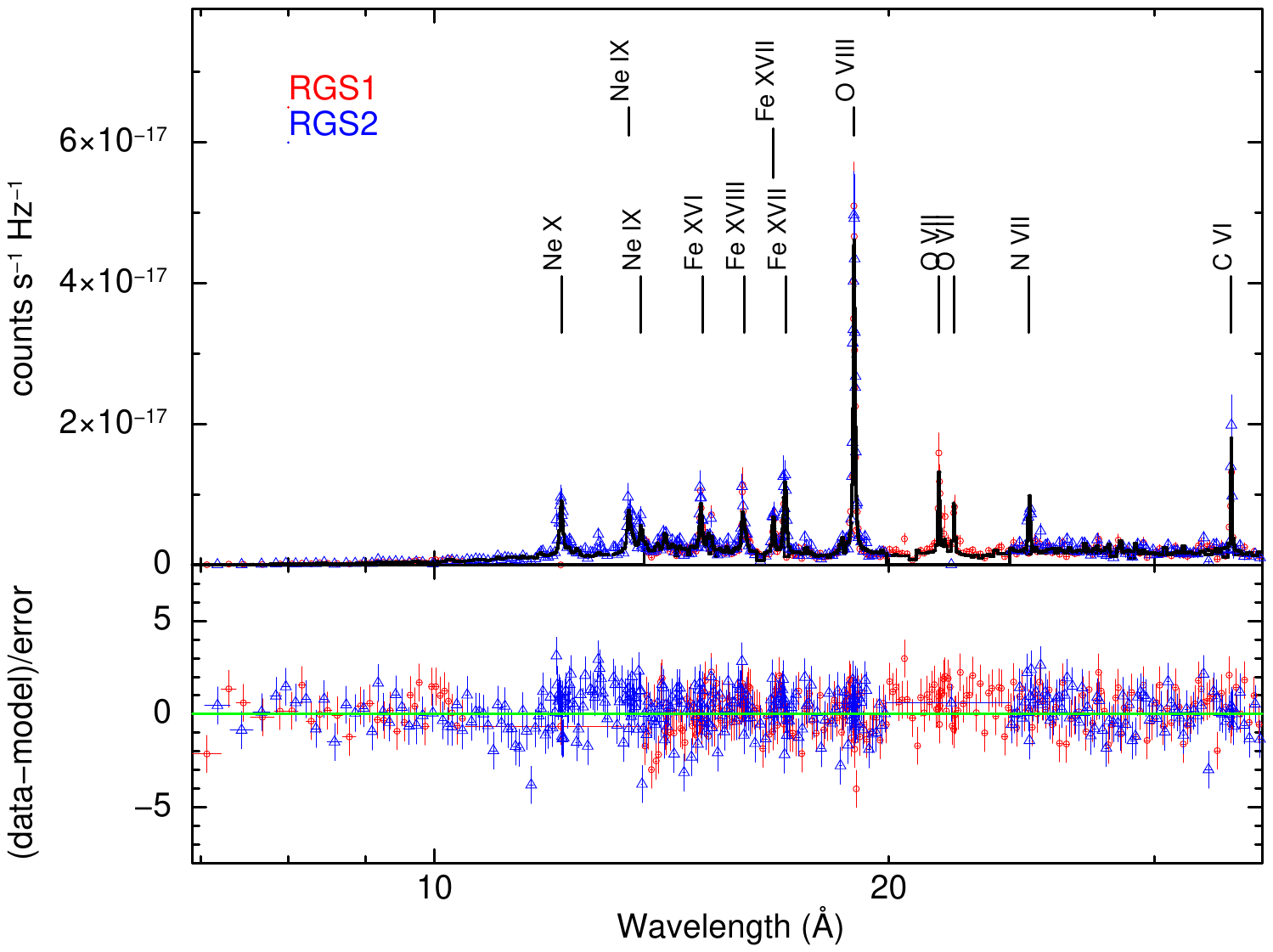}}\label{fig:rgs_lqhya2}
    
    \caption{RGS1 (in red) and RGS2 (in blue) spectra with the best fit {\sc 3T-vapec} model (in black). The lower panels show the residuals of the fit. The prominent emission lines are also marked.}
    \label{fig:RGSspectra}
\end{figure*}

\section{RGS Spectral analysis}\label{sec:RGS}

The analysis of EPIC-PN data revealed that the global abundance changes are confined to a fraction of the total observation period, with variations not exceeding twice the quiescent state abundances for all three stars. Consequently, when the entire observation is taken into account, the abundance variations that occur during the flare tend to be averaged. To quantify the elemental abundances for each observation, we simultaneously fitted the RGS1 and RGS2 spectra using an absorbed three-temperature plasma model ({\sc vapec} in XSPEC). 
Similar to EPIC-PN analysis, $N_{\text{H}}$ was fixed to 10$^{18}$ atom cm$^{-2}$. All temperature and emission-measure components of the model were allowed to vary during the fit. The coronal abundances of each model component were tied and allowed to vary, with the exception of He, Al, Ca, and Ni, which were fixed to solar photospheric values given by \cite{1989GeCoA..53..197A}. The best fit spectra for each observation are presented in Figure \ref{fig:RGSspectra}. Here, in each Figure, the upper panels show the RGS1 and RGS2 spectra along with the best fit 3T-{\sc vapec} model. In the lower panel, the residuals for each fit are shown. The prominent atomic lines were also marked in the RGS spectra. The Table \ref{tab:rgs} shows the best-fit spectral parameters.

\begin{deluxetable*}{lcccc}
\tablecaption{Spectral parameters obtained by the best fit {\sc 3T-vapec} model over the RGS spctra for all three stars.}
\label{tab:rgs}
\tablewidth{0pt} 
\tablehead{
    \colhead{Parameters($\downarrow$) / Stars($\rightarrow)$} & \colhead{V834 Tau} & \colhead{BY Dra} & \colhead{LQ Hya (epoch-I)} & \colhead{LQ Hya (epoch-II)} 
}
\startdata
kT1 (keV) & $0.27_{-0.01}^{+0.01}$ & $0.26_{-0.01}^{+0.01}$ & $0.26_{-0.01}^{+0.01}$ & $0.27_{-0.01}^{+0.01}$ \\
kT2 (keV) & $0.69_{-0.02}^{+0.02}$ & $0.66_{-0.02}^{+0.02}$ & $0.66_{-0.02}^{+0.02}$ & $0.71_{-0.02}^{+0.02}$ \\
kT3 (keV) & $1.33_{-0.10}^{+0.13}$ & $1.67_{-0.23}^{+0.37}$ & $1.41_{-0.08}^{+0.10}$ & $1.49_{-0.10}^{+0.12}$ \\
EM1 ($10^{52}$ cm$^{-3}$) & $0.81_{-0.07}^{+0.07}$ & $0.41_{-0.06}^{+0.07}$ & $1.23_{-0.13}^{+0.13}$ & $1.38_{-0.10}^{+0.10}$ \\
EM2 ($10^{52}$ cm$^{-3}$) & $1.25_{-0.11}^{+0.11}$ & $0.53_{-0.07}^{+0.08}$ & $1.92_{-0.17}^{+0.19}$ & $1.89_{-0.12}^{+0.13}$ \\
EM3 ($10^{52}$ cm$^{-3}$) & $0.80_{-0.09}^{+0.08}$ & $0.39_{-0.07}^{+0.07}$ & $2.40_{-0.15}^{+0.15}$ & $1.66_{-0.12}^{+0.11}$ \\
C (Z/Z$_\odot$) & $1.87_{-0.20}^{+0.22}$ & $0.92_{-0.21}^{+0.25}$ & $0.83_{-0.11}^{+0.12}$ & $0.66_{-0.08}^{+0.08}$ \\
N (Z/Z$_\odot$) & $1.74_{-0.18}^{+0.20}$ & $0.45_{-0.12}^{+0.14}$ & $0.55_{-0.09}^{+0.09}$ & $0.66_{-0.07}^{+0.08}$ \\
O (Z/Z$_\odot$) & $0.72_{-0.05}^{+0.06}$ & $0.43_{-0.05}^{+0.06}$ & $0.41_{-0.03}^{+0.04}$ & $0.37_{-0.02}^{+0.02}$ \\
Ne (Z/Z$_\odot$) & $2.41_{-0.19}^{+0.22}$ & $1.64_{-0.20}^{+0.24}$ & $1.21_{-0.10}^{+0.12}$ & $1.32_{-0.09}^{+0.09}$ \\
Mg (Z/Z$_\odot$) & $0.47_{-0.09}^{+0.09}$ & $0.40_{-0.15}^{+0.15}$ & $0.31_{-0.08}^{+0.08}$ & $<0.74$ \\
Si (Z/Z$_\odot$) & $<1.42$ & $2.47_{-1.19}^{+1.30}$ & $0.76_{-0.57}^{+0.59}$ & $<0.73$ \\
S (Z/Z$_\odot$) & $0.82_{-0.19}^{+0.21}$ & $0.47_{-0.28}^{+0.31}$ & $0.27_{-0.13}^{+0.14}$ & $0.45_{-0.10}^{+0.11}$ \\
Ar (Z/Z$_\odot$) & $1.15_{-0.26}^{+0.29}$ & $0.96_{-0.37}^{+0.44}$ & $0.81_{-0.21}^{+0.24}$ & $0.92_{-0.15}^{+0.16}$ \\
Fe (Z/Z$_\odot$) & $0.26_{-0.02}^{+0.03}$ & $0.21_{-0.03}^{+0.03}$ & $0.18_{-0.02}^{+0.02}$ & $0.19_{-0.01}^{+0.01}$ \\
$\chi^2_{red}/dof$&1.02/315&1.36/798&1.16/886&1.23/607\\
\enddata
\tablenotetext{}{The abundances are relative to solar photospheric abundances from \cite{1989GeCoA..53..197A}. The error bars are set at 68\% confidence limit.}
\end{deluxetable*}

\subsection{FIP and inverse-FIP effect}
Variations in elemental abundances between stellar coronae and their underlying photospheres are commonly attributed to fractionation processes linked to the first ionization potential (FIP). In slowly rotating stars such as the Sun, elements with FIP values below $\sim$10 eV are typically enhanced in the corona relative to high-FIP elements. Such enhancement of low FIP elements in corona is known as the FIP effect \citep[][]{1992PhyS...46..202F,2000PhyS...61..222F}. In some rapidly rotating active stars, the trend is reversed, with high-FIP elements enhanced, resulting in an inverse-FIP effect \citep[for a detailed review of FIP/inverse-FIP see][]{2015LRSP...12....2L}. However,  some stars show no discernible FIP bias in their coronal abundances \citep[][]{2004A&A...416..281S, 2006ApJ...643..444W}.

To check if there is any fractionation bias in the coronae of these stars, we first look for photospheric iron abundances and compare them with the coronal iron abundances obtained from {\sc vapec} fits. The photospheric [Fe/H] abundance of BY Dra is estimated to be +0.2 by \cite{2012A&A...547A.106M}. For V834 Tau and LQ Hya the [Fe/H] values are 0.11$\pm$0.07 and 0.28$\pm$0.1, respectively \citep[][]{2022A&A...663A...4S}. Using these photospheric values, the fractionation of iron can be directly probed by taking the ratio of coronal iron with respect to the solar to the photospheric iron. The Fe$_{corona}$/Fe$_{photosphere}$ ratio for stars V834 Tau, BY Dra, and LQ Hya was found to be  0.20$\pm$0.04, 0.13$\pm$0.02, and 0.10$\pm$0.02, respectively. This implies that iron abundances were depleted by a factor of $\sim$5-10 relative to the respective stellar photospheric values, suggesting an inverse-FIP effect akin to fractionation. 

\begin{figure*}
    \centering
    \includegraphics[width=0.95\textwidth]{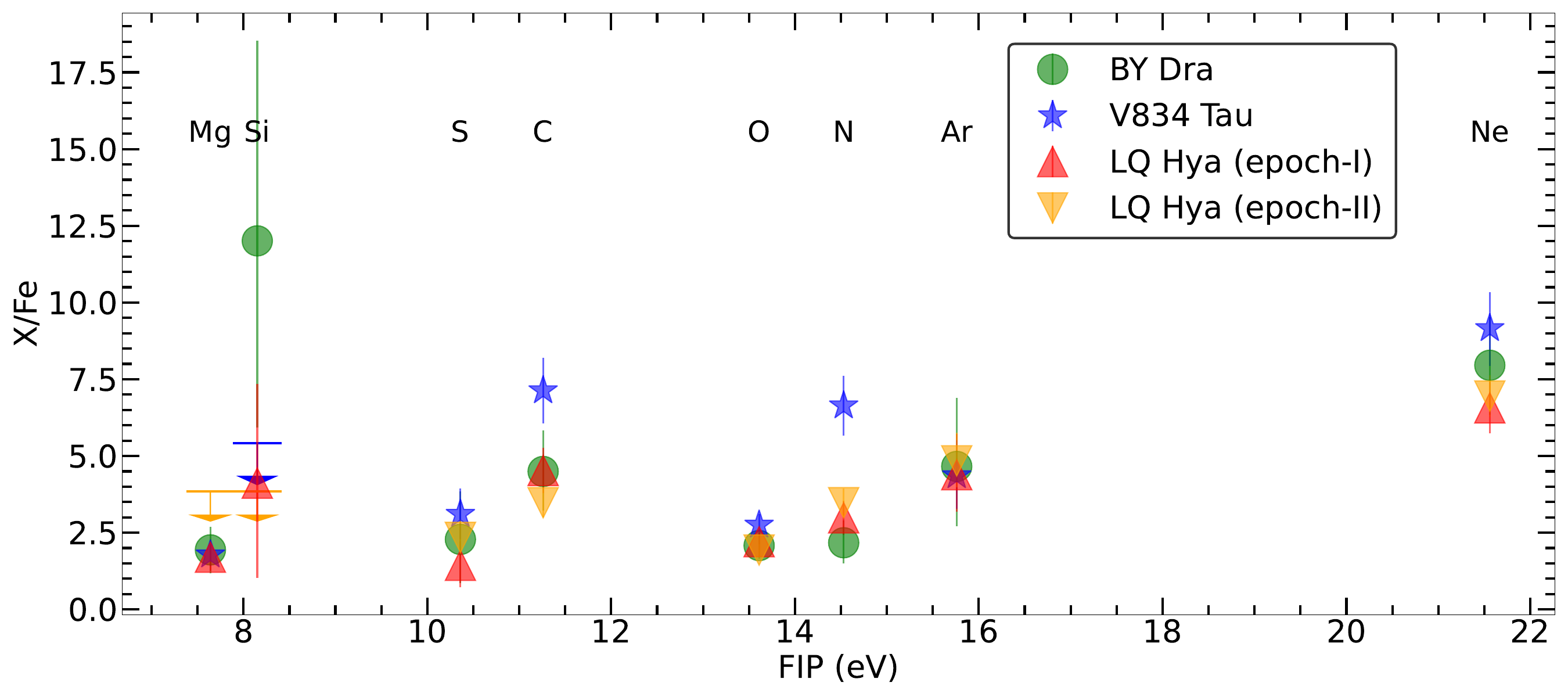}
    \caption{Plot of elemental abundances relative to Fe against the FIP.}
    \label{fig:FIP}
\end{figure*}

Since photospheric abundances for different elements were unavailable, we normalized coronal elemental abundances to iron and plotted them against the corresponding FIP values.
The  Figure \ref{fig:FIP} shows normalized elemental abundances vs FIP value. We found that all three stars show similar trends in their coronal abundances against FIP. The high FIP element, Ne, shows the greatest enhancement in Fe abundances.
The Ne/Fe ratio for V834 Tau and BY Dra were found to be  9.2$\pm$1.2 and 8.0$\pm$0.7, respectively, whereas for epoch-I and epoch-II observations of LQ Hya, the Ne/Fe ratios were found to be 6.6$\pm$0.8 and 6.9$\pm$0.7, respectively. The depletion of coronal Fe relative to photospheric values, together with the enhancement of the high-FIP element Ne relative to coronal Fe, indicates an inverse FIP effect in all three stars.

\section{Flare loop modeling}\label{sec:loop}
 
 We derive the loop length of the flaring plasma using the state-of-the-art loop model of \cite{1997A&A...325..782R}. This model incorporates the effect of significant heating during flare decay, utilising a time-dependent hydrodynamic loop model. The model estimates the semi-loop length as
\begin{equation}
{\rm L = 2.7\times10^3\frac{\tau_{d}T_{max}^{1/2}}{F(\zeta)}  ~~cm~~
~~ for ~~~0.35 < \zeta \leq 1.6}
\label{eq:loop}
\end{equation}
The parameters  $F(\zeta)$ and maximum flare temperature (T$_{max}$) were calculated using the following equations 
\begin{equation}
    T_{max} = 0.13 T_{obs}^{1.16},
    \quad F(\zeta) = \frac{0.51}{\zeta-0.35}+1.36
    \label{eq:tmax}
\end{equation}

Here, $T_{obs}$ represents the peak observed temperature, and $\zeta$ is the slope of the density-temperature diagram. The factor 1/$F(\zeta)$ is termed the correction factor to the loop length estimation by \cite{seriomodel1991}, whose estimates are based on a freely decaying loop model.
The correction factor  1/$F(\zeta)$ ranges from $\sim$ 0.02 for strong sustained heating ($\zeta$=0.36) and  $\sim$0.57 no sustained heating ($\zeta$=1.6), respectively. So, the \cite{seriomodel1991} model overestimates the loop length by 98 to 43 \% for strong heating to no heating.  We took the square root of the emission measure ($\sqrt{EM} $) as a proxy of density and derived   $\zeta$ using the $\sqrt{EM}$ -- T curve. 

The pressure of the plasma inside the loop can be calculated from  Rosner–Tucker–Vaiana (RTV) scaling laws as p = (T/1400)$^3$/L (dyne/cm$^2$).  The RTV scaling laws apply near the flare peak, where the coronal loop is approximately in steady-state hydrostatic and thermal equilibrium.
Assuming the fully ionised plasma with 10\%  of He, the density can be estimated as  $n_e$=p/(1.8k$_B$T$_{max}$) cm$^{-3}$, where k$_B$ is the Boltzman constant. 
By combining the density and peak emission measure of the flare, flaring volume can be estimated as V=EM$_{max}$/0.8$n_e^2$.
 Considering the maximum flare temperature, derived loop lengths, rise and decay times, the heating rate per unit volume ($HR_V$) and the  total energy ($E_{tot}$) corresponding to the heating rate were estimated using RTV scaling laws as 
 $HR_V=T_{max}^{7/2}/(10^6L^2)$, and $E_{tot}= HR_VV(\tau_r+\tau_d)$, respectively. The $E_{tot}$ is an estimation for the total energy deposited in the coronal loop during the flare.
Based on plasma pressure, loop volume, and total energy, the minimum magnetic field ($B_{min}$) required to constrain the flaring plasma in the coronal loop and the total magnetic field ($B_{tot}$) that gives rise to flaring can be estimated by using a simple pressure balance equation in magnetohydrodynamics, as 
$B_{min} = \sqrt{8\pi p} \quad (G)$ and  $B_{tot}= \sqrt{8\pi E_{tot}/V+B{min}^2}$ \quad (G), respectively. The derived loop parameters are given in Table \ref{tab:loopparam} and described below.

\subsection{V834 Tau}
For the flare F1 from V834 Tau, we found the value of $\zeta$ to be 0.69$\pm$0.38. Considering this value, the loop length was estimated to be 1.1$\pm$0.6$\times$10$^{10}$ cm. The maximum flare temperature using equation (2) was derived to be 62$\pm$8 MK. The pressure and density for the flare F1 were found to be 7.9$\pm$5.3$\times10^3$ dyne and 5.1$\pm$3.5$\times10^{11}$cm$^{-3}$, respectively. Considering the lower value of density and upper value of maximum emission measure, the upper limit on volume of flaring region was derived to be $<8 \times10^{29}$ cm$^3$. We derived the upper limit on the heating rate per unit volume using the upper limit on $T_{max}$ and the lower limit on loop length. The resulting upper limit on $HR_V$ was estimated to be 115 erg s$^{-1}$ cm$^{-3}$. By using the upper limits of $HR_V$ and volume of the flaring region, the upper limit on the total energy for the flare was estimated 2$\times10^{35}$ erg, which is 227 times higher than the total X-ray energy emitted. We found the minimum magnetic field required to trap such a plasma should be of strength 450$\pm$160 G and the total magnetic field should be more than 2.8 kG.

\subsection{BY Dra}
The maximum flare temperature for flare F2 was found to be 63$\pm$20 MK. The value of $\zeta$ was found to be  2.18$\pm$1.31. Considering the lower limit of the $\zeta$ (0.87), we have estimated the lower bound to the flaring loop length as 1.3$\times10^{10}$ cm.  The value of density and pressure during the flare F2 should be more than 1.1$\times10^{11}$ cm$^{-3}$ and 2.2$\times10^3$ dyne, respectively. Based on the lower bounds on density and loop length; the upper limits of volume, $HR_V$, and $E_{tot}$ were estimated, which come out to be less than 1.6$\times10^{30}$ cm$^3$, 31 erg s$^{-1}$ cm$^{-3}$, and 10$^{35}$ erg, respectively. The $E_{tot}$ is $\sim$100 times higher than the total energy emitted in X-rays for flare F2. We find that the minimum magnetic and total magnetic fields for flare F2 should be more than 235 G and 1.3 kG, respectively.

\subsection{LQ Hya}
For the flares F3, F4, and F6, the values of $\zeta$ were found to be beyond the applicability limits of \cite{1997A&A...325..782R} model; thus, we used \cite{seriomodel1991} model in order to estimate the loop length of flaring plasma. For the flare F5, the value of $\zeta$ is equal to 5.41$\pm$4.61, thus we considered the lower limit to $\zeta$ and proceeded with the model of  \cite{1997A&A...325..782R} to derive loop length. The loop length for flares F3, F4, and F6 were found to be 10$\times10^{10}$, 3$\times10^{10}$, and 14$\times10^{10}$ cm, respectively. Whereas for the flare F5, we provide the lower limit of the loop length as  2.3$\times10^{10}$ cm. We found that the plasma pressure for all the flares in LQ Hya was less than 4.7$\times10^4$ dyne, whereas the density was less than 6$\times10^{12}$ cm$^{-3}$. The volume of flaring regions for all flares from LQ Hya was estimated in the range of  10$^{26-30}$ cm$^3$. We found heating rate per unit volume for flare F3 to be $\sim$1.3 erg s$^{-1}$ cm$^{-3}$, whereas for flare F4, F5, and F6 the upper limit on $HR_V$ was found out to be 87, 7 and 0.33 erg s$^{-1}$ cm$^{-3}$, respectively. The total energy due to heating rate could be estimated for flare F3 as 1.4$\times10^{35}$ erg, which was $\sim$33 times higher than the X-ray energy during the flare. For the flares F4, F5, and F6, the total heating energy constraints were not feasible due to the upper and lower bounds on $HR_V$ and the volume. We find the minimum magnetic field for the flares F3 - F6 to be less than 1 kG. The total magnetic field for flare F3 was estimated to be more than 765 G whereas for other flares (F4-F6) the total magnetic field estimates were not possible. 

 It is important to note that the loop lengths for the flares F3, F4, and F6 were calculated based on the \cite{seriomodel1991} model. Consequently, the loop length estimates for flares F3, F4, and F6 were likely overestimated by up to a factor of 2, as discussed earlier. As a result, the pressure and density would be underestimated by a factor of 2, whereas the volume would be overestimated by a factor of 4. Similarly, the heating rate per unit volume would be underestimated by a factor of 4, whereas the E$_{tot}$ would remain the same in both models.
Whereas the minimum magnetic field would be underestimated by a factor of $\sqrt{2}$.

\begin{deluxetable*}{lcccccc}
\tablecaption{Loop parameters of analysed flares.}
\label{tab:loopparam}
\tablewidth{0pt} 
\tablehead{
    \colhead{Parameters($\downarrow$) / Flares($\rightarrow)$} & \colhead{F1} & \colhead{F2} & \colhead{F3} & \colhead{F4} & \colhead{F5} & \colhead{F6} 
}
\startdata
Rise time ($\tau_r$) (s)                        &   788$\pm$108         & \ldots        & 2146$\pm$166      & 660$\pm$179       & \ldots             & 247$\pm$52       \\
Decay time ($\tau_d$) (s)                       &   1486$\pm$162        & 1916$\pm$137  & 3649$\pm$284      & 1198.$\pm$325      & 4048$\pm$399       & 4886$\pm$948     \\
Maximum Temperature (T$_{max}$) (MK)            &   62$\pm$8            & 63$\pm$20     & 109$\pm$24        & 80$\pm$50         & 56$\pm$20          & 54$\pm$19        \\
$\zeta$                                         &   0.69$\pm$0.38       & $>$0.87       & --                & --                & $>$0.8             & --               \\
Total energy (E$_{X}$) ($10^{33}$ erg)          &   0.88$\pm$0.02       & 1.03$\pm$0.02 & 4.23$\pm$0.04     & 0.57$\pm$0.02     & 1.4$\pm$0.02       & 2.56$\pm$0.03    \\
Loop Length (L) ($10^{10}$ cm)                  &   1.1$\pm$0.6   & $>$1.3        & *10.02$\pm$1.35   & *2.9$\pm$1.2      & $>$2.3             & *14$\pm$4        \\
Pressure (p) ($10^3$ dyne)                      &   7.9$\pm$5.3         & $>$2.2        & 4.7$\pm$3.2       & $<$47             & $<$6.9             & $<$1.4           \\
Density (n$_e$) ($10^{11}$ cm$^{-3}$)           &   5.1$\pm$3.5           & $>$1.1         & 1.7$\pm$1.2         & $<$63.1            & $<$7.7              & $<$1.6            \\
Loop Volume (V) ($10^{30}$ cm$^3$)              &   $<$0.8              & $<$1.6        & $<$9.4            & $>$0.0003         & $>$0.03            & $>$1.3           \\
Heating rate (HR$_V$) (erg s$^{-1}$cm$^{-3}$)   &   $<$115              & $<$30.8       & 1.3$\pm$1.1       & $<$87             & $<$7               & $<$0.33          \\
Total Energy (E$_{tot}$) ($10^{33}$ erg) &   $<$234              & $<$101        & $<$140            & --                & --                 & --               \\
Minimum magnetic field (B$_{min}$) (G)          &   450$\pm$160         & $>$235        & 340$\pm$120       & $<$1087           & $<$416             & $<$188           \\
Total magnetic field (B$_{tot}$) (G)            &   $>$2779             & $>$1281       & $>$765            & --                & --                 & --               \\
GOES class        & X190           &X224 & X285 &X278&X197&X432\\
\enddata
\tablenotetext{}{* Parameter estimated using the loop model of \cite{seriomodel1991}.}
\end{deluxetable*}

\section{Discussion}\label{sec:discuss}
\subsection{Temporal properties of flaring and quiescent states}
We investigated the X-ray quiescent and flaring states of three K-type main-sequence active stars: V834 Tau, BY Dra, and LQ Hya. The quiescent X-ray luminosity for stars V834 Tau, BY Dra, and LQ Hya is found to be similar to that of other active late-type stars \citep[][]{1999ApJ...512..874S,2008MNRAS.387.1627P}. The peak flare X-ray luminosities for flares from V834 Tau, BY Dra, and LQ Hya were found to be $\sim$2.2, $\sim$1.5, and 1.3 -- 1.5 times to that of the quiescent state X-ray luminosity, respectively, suggesting that these flares are as strong as observed in several other active stars \citep[e.g.][]{2008MNRAS.387.1627P,2015A&A...581A..28P,2024ApJ...961..130Z}. 
However, the increment in X-ray luminosity has been observed $\ge$ 100 times the quiescent value for some stars, e.g. AB Dor \citep[][]{2000A&A...356..627M}, Proxima Cen \citep[][]{2004A&A...416..713G}, and LMC 335 \citep[][]{2012ApJ...754..107T}.
Moreover, the star EV Lac showed the strongest flare to quiescent enhancement, where the peak flare luminosity was found to be $\sim$7000 times that of the quiescent state luminosity \citep[][]{2010ApJ...721..785O}.

For the majority of the flares, we found the rise times to be $\sim$1.8 times shorter than the decay times, exhibiting the standard fast rise and slow decay,  consistent with observations in other stellar flares from late-type stars. However, for the flare F6, the rise time was $\sim$300 sec while the decay time was  $\sim$5 ks. There are not many flares with such a very-fast rise and slower decay flare profile that are reported 
\citep[e.g. ][]{2015A&A...581A..28P,2024ApJ...961..130Z}. Considering the flare F6 to be similar to other flares studied, its decay time should be $\sim$600 sec. This expected flare decay is 8 times shorter than the observed value. The reason for such a discrepancy might be the presence of unresolved flares during the decay phase or a very high heating rate, which slows the decay process. We will discuss flare F6 later in the context of the time-resolved spectral analysis to understand its likely origin.

\subsection{The quiescent emission: temperatures, emission measures and abundances}
The quiescence of all three stars was well explained by a two-temperature plasma model, with the cooler component around 0.26--0.3 keV and the hotter component near 0.93--1.01 keV.
 Despite the consistency in temperature across both thermal components, the corresponding emission measures differ significantly among these stars. The cool component dominates the emission in V834 Tau, the hot component dominates in LQ Hya, and both components contribute comparably in BY Dra. It is worth noting that V834 Tau and LQ Hya are both $\sim$50 Myr old and have similar spectral types, yet their coronal properties differ substantially. The divergent coronal behavior of these stars appears primarily driven by their respective rotation periods. For instance, V834 Tau rotates at half the rate of LQ Hya. However, BY Dra, whose rotation period lies between the two, exhibits nearly equal emission measures for its cool and hot components.  It appears that there exists a transitional rotation period during which the dominant emission measure shifts between the cool and hot coronal components.  However, future studies involving larger samples will be crucial for reaching a definitive conclusion.

During quiescence, the global abundances are approximately 0.1–0.2 times the solar photospheric values,  a level of coronal underabundance commonly observed in stars. When compared with results from high-resolution spectral fits, the global abundances derived using the {\sc apec} model were found to be broadly consistent with the Fe abundances obtained from the {\sc vapec} model \textbf{(see Section \ref{sec:RGS})}. This suggests that global abundances are primarily influenced by iron abundance, which in turn dominates because of strong Fe emission lines in the soft X-ray spectral range.

\begin{figure}
    \centering
    \includegraphics[width=\columnwidth]{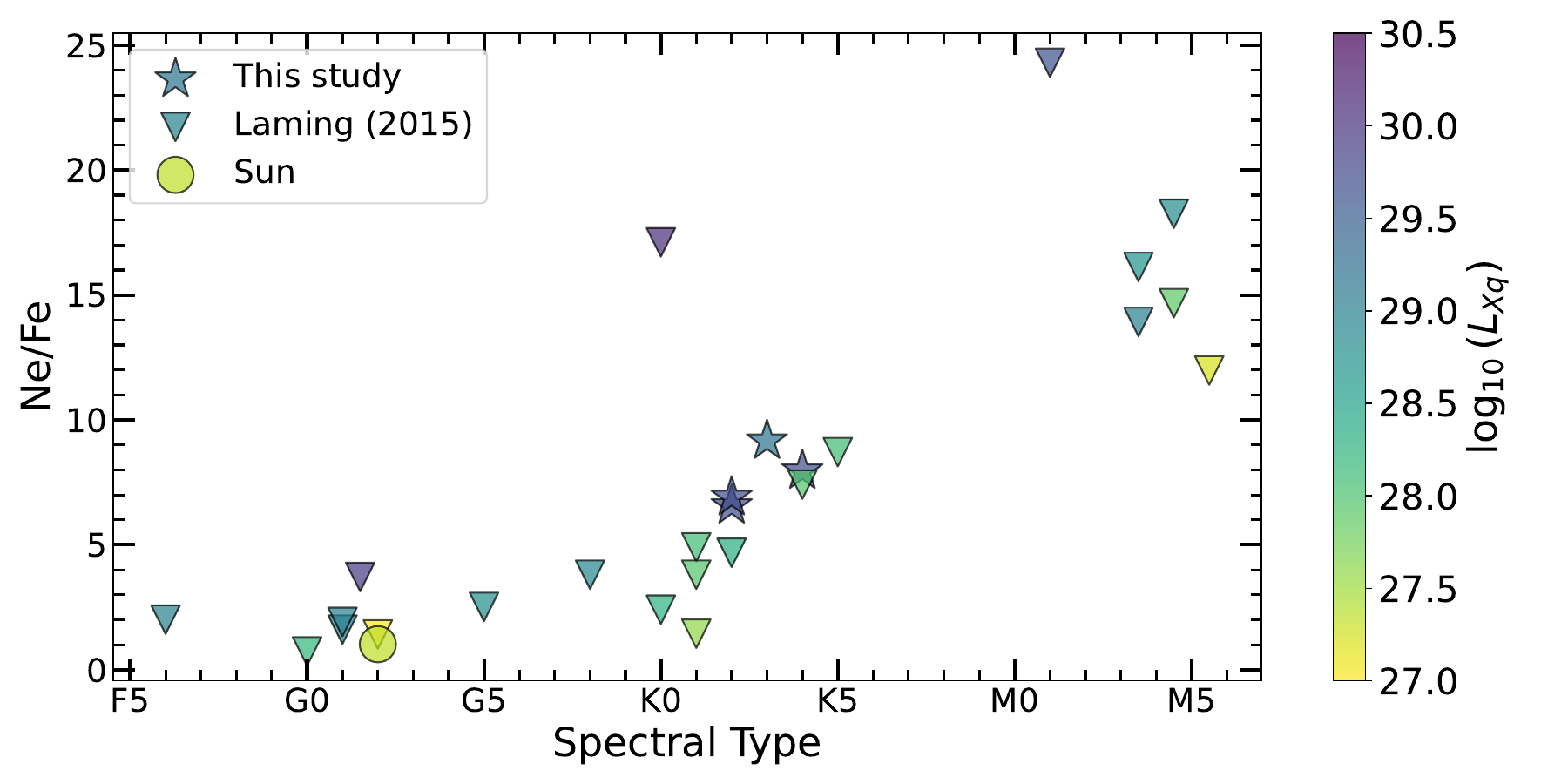}
    \caption{Ne/Fe as a function of the spectral class. The Ne/Fe values for other stars are taken from  \cite{2015LRSP...12....2L}. The Sun is denoted by a circle.}
    \label{fig:fipstarsandsun}
\end{figure}

The coronal elemental abundances provide insights into the fractionation processes in the coronae. The depletion of Fe abundance relative to the stellar photosphere and the trends in coronal abundance with FIP indicate that the stars in the current study exhibit the inverse-FIP effect. 
We plotted the Ne (high FIP element) to Fe (low FIP element) ratio  (Ne/Fe) as a function of spectral class of the active stars in the Figure \ref{fig:fipstarsandsun}. We used data from Table 2 of \cite{2015LRSP...12....2L} for the Ne/Fe ratio of other active stars and the Sun. We also used a color bar to indicate the X-ray quiescent luminosity in the same figure. The Ne/Fe for the Sun is also plotted as a circle for comparison. 
The Ne/Fe ratios of V834 Tau, BY Dra, and LQ Hya were consistent with the typical range observed in K-type stars. 
We also noted that within a given spectral class, the stars with the highest Ne/Fe ratio have higher quiescent X-ray luminosity, indicating a stronger magnetic dynamo and a stronger fractionation bias. This is the opposite of what solar observations suggest: at solar maxima, the lower FIP elements get stronger enhancement, while higher FIP elements get suppressed more. 
This indicates that stellar coronae exhibit inherently different abundance patterns from those observed in the solar corona.

\subsection{Spectral evolution of flare}
 
In general, we found that the flare temperature peaks before the flare emission measure and X-ray luminosity, whereas the flare abundances peak in the early decay phase. This trend in plasma parameters is consistent with the standard flare model \citep[see][for a detailed discussion]{2007A&A...471..271R}. However, there are a few cases in which both temperature and emission measure peaked simultaneously, likely due to the coarse time resolution of spectral analysis. The flare evolution can be summarized as follows. A heat pulse during the pre-flare/rise phase rapidly heats the loop apex, leading to a temperature peak. This heat then reaches footpoints, leading to the evaporation of cooler chromospheric plasma into the loop, increasing its density. After the heat pulse stops, the loop cools by thermal conduction, but evaporation continues to increase density until radiative and conductive losses balance, stopping evaporation and causing density to decay. Consequently, temperature peaks before peak emission, and coronal abundances peak after peak luminosity.

BY Dra, the prototype of a variable star, showed an X-ray flare with an energy of 1.03$\times$10$^{33}$ erg, placing it in the superflare category. Based on the ephemeris
HJD = 2453999.7144(21) + 5.9751130(46) E,  the flare peak occurred at an orbital phase of 0.1390$\pm$0.0006, indicating that the two binary components were near the periastron passage. Here, HJD is the heliocentric Julian Day, and the epoch of zero phase is HJD of the closest approach between the binary components \citep{2012mnras.419.1285h}.   The binary separation at phase 0.14  of  $\sim$15 R$_\odot$  is $\sim$115 times larger than the loop length estimate. A single flaring episode has been reported in the past, occurring on 24 September 1984 \citep{1986A&A...156...95D}, and this flare was also triggered at an orbital phase  $\sim$0.14. The occurrence of flaring episodes near the same phase, close to periastron passage, in two different epochs suggests that the origin of flares cannot be ruled out from inter-binary magnetic reconnection. \citep[][]{1983ASSL..102..629U}. 

 Previously, an X-ray flare from LQ Hya was studied by \cite{2001A&A...371..973C} using ASCA and ROSAT data, which yield similar loop parameters to those obtained for the four flares observed in the current study. However, the flares analyzed in this study have energy about an order of magnitude lower than that observed in the earlier flare. The long decay time of flare F6 is likely due to either unresolved flares, reheating within a single loop, or a flare within a loop arcade. We notice that the temperature and emission measures attain similar values during the later phases of flare F6, and the entire flare is not captured during the observation. This indicates that the flare F6 is likely to have originated in a loop arcade, or that the loop was reheated during the later phases. However, the incomplete coverage of the flare’s evolution precludes a definitive conclusion.
  
 Using the ephemeris HJD = 2445274.220(13) + 1.601136(13) E, where epoch corresponds to 0 phase, represents the HJD of the first light-curve minimum \citep{1993A&A...276..345J}, the flares F3 and F4 reached their peak emission at rotational phases 0.90 $\pm$ 0.04 and 0.95 $\pm$ 0.04, respectively. While the flares F5 and F6 peaked at phases 0.86$\pm$0.04 and 0.95$\pm$0.04, respectively. The remarkable phase similarity for flares F4 and F6 of LQ Hya suggests that the active region around this phase was complex, resulting in flaring events, even though the time difference between the two flares was about 6 months. This directly indicates that during 2003, LQ Hya possessed a strong magnetic region around phase 0.95 that remained stable for at least six months. 
  The flares in the current study occurred around this phase, after more than a decade of optical observations, indicating persistent starspots near phase $\sim$0 in LQ Hya.  

The global abundances during the flare F1 from V834 Tau were enhanced by a factor of up to 1.47; whereas for other flares, we could not find any significant enhancement.  The enhancement of coronal abundances during the flare has also been reported in the past for active stars \citep[][]{2001A&A...365L.336G,2008A&A...482..639N,2010A&A...514A..94L,2024MNRAS.527.1705D,2025NewA..11402295S}, which were linked to the chromospheric evaporation. 

\begin{figure*}
\centering
\subfigure[]{\includegraphics[width=0.95\columnwidth]{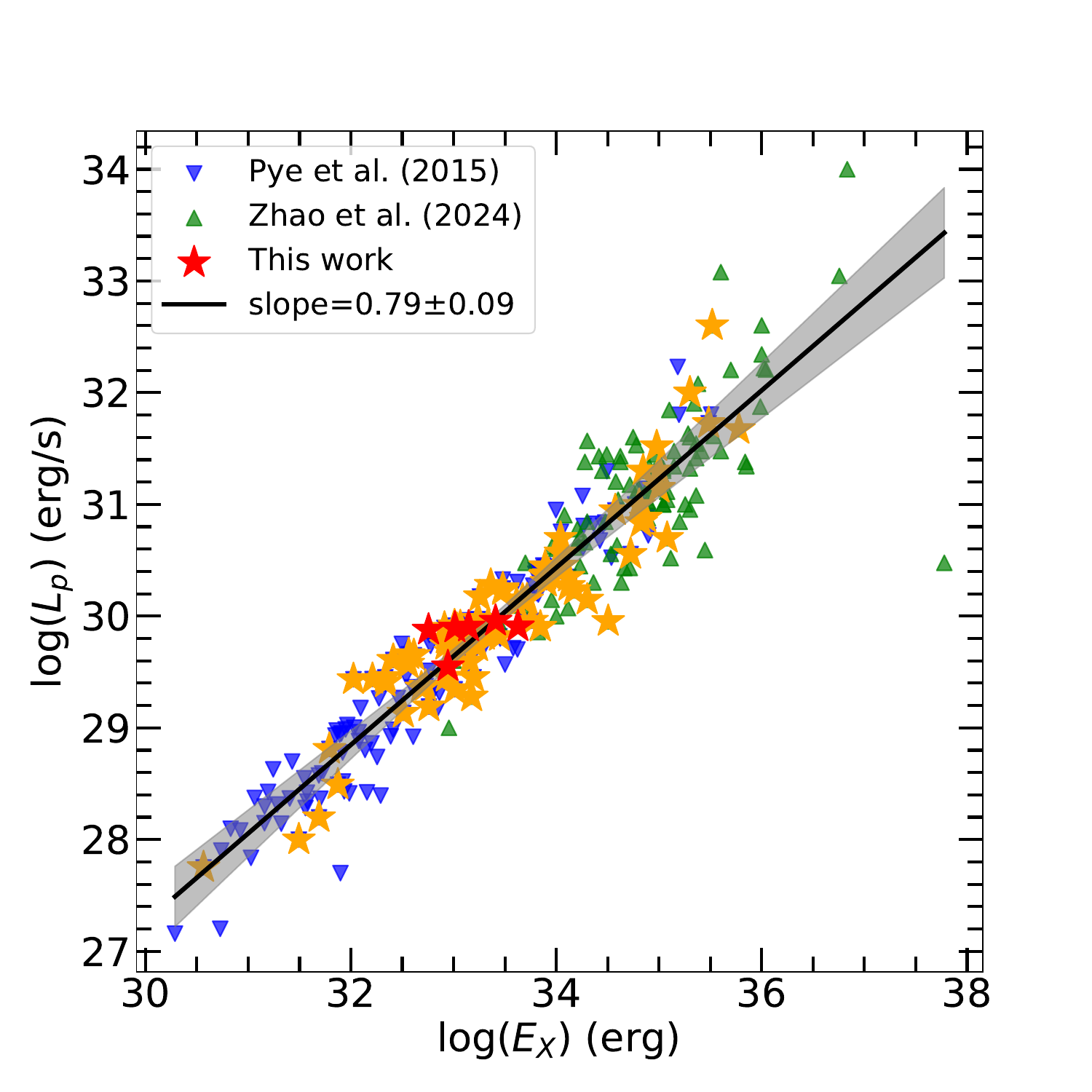}}
\subfigure[]{\includegraphics[width=0.95\columnwidth]{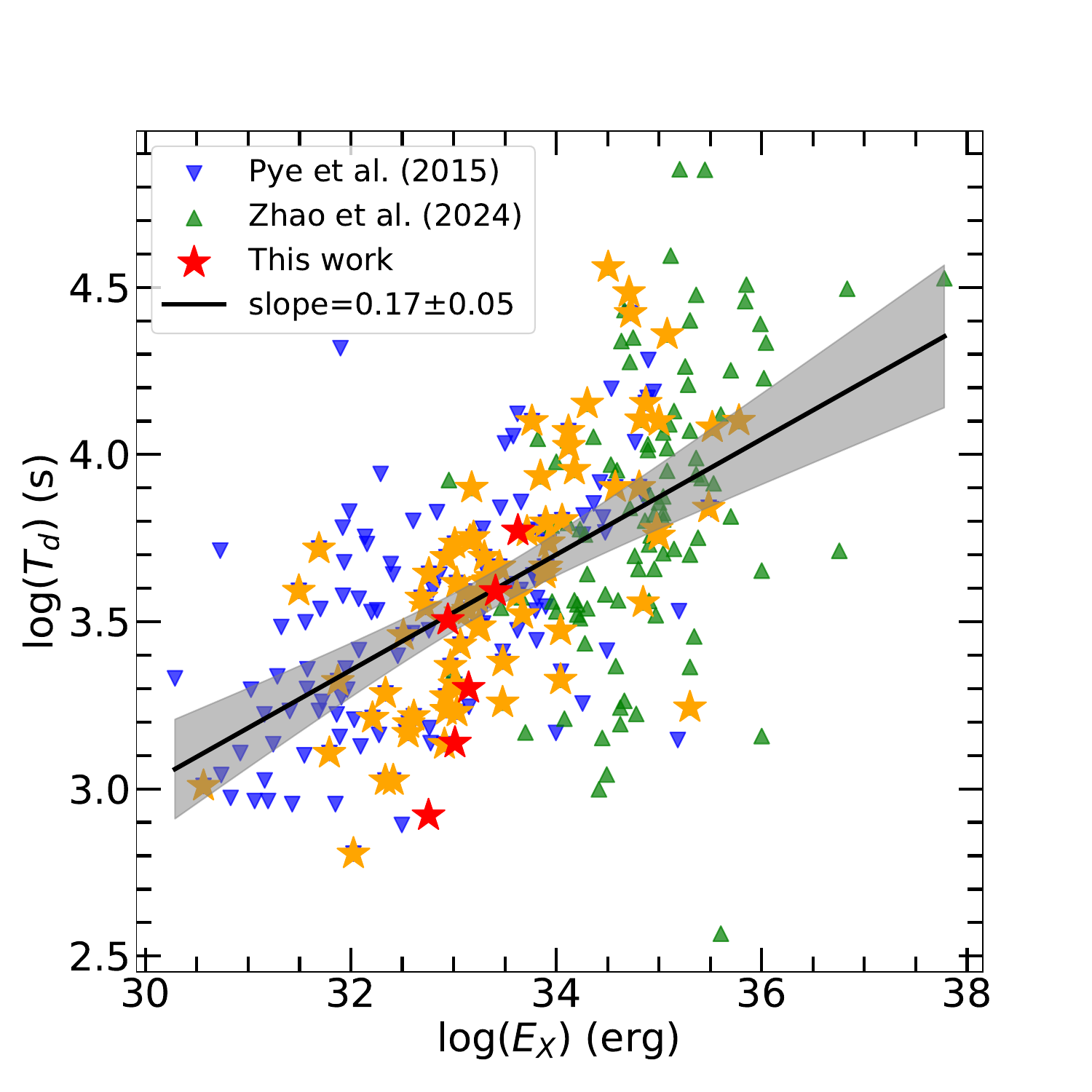}}\\
\subfigure[]{\includegraphics[width=0.95\columnwidth]{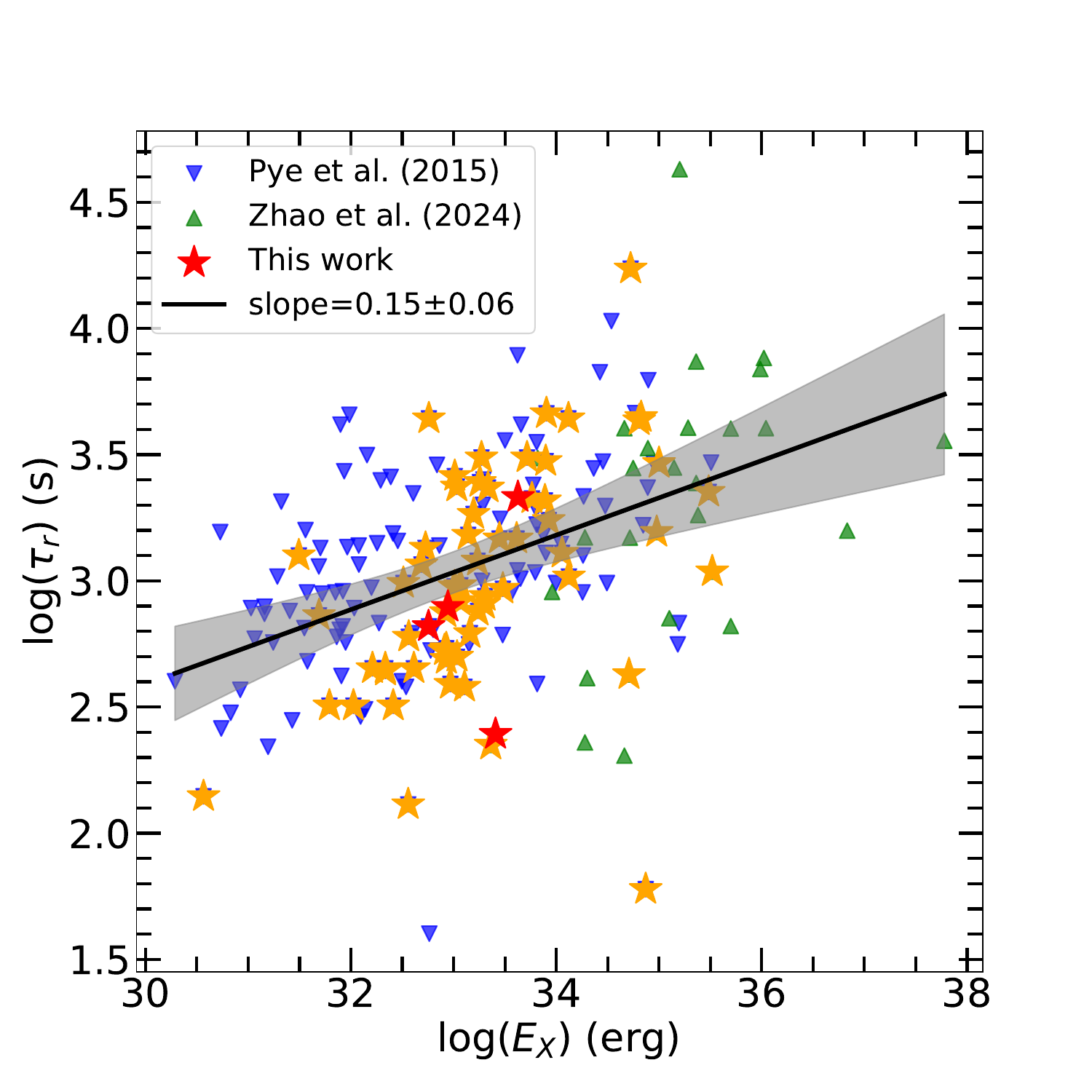}}
\subfigure[]{\includegraphics[width=0.95\columnwidth]{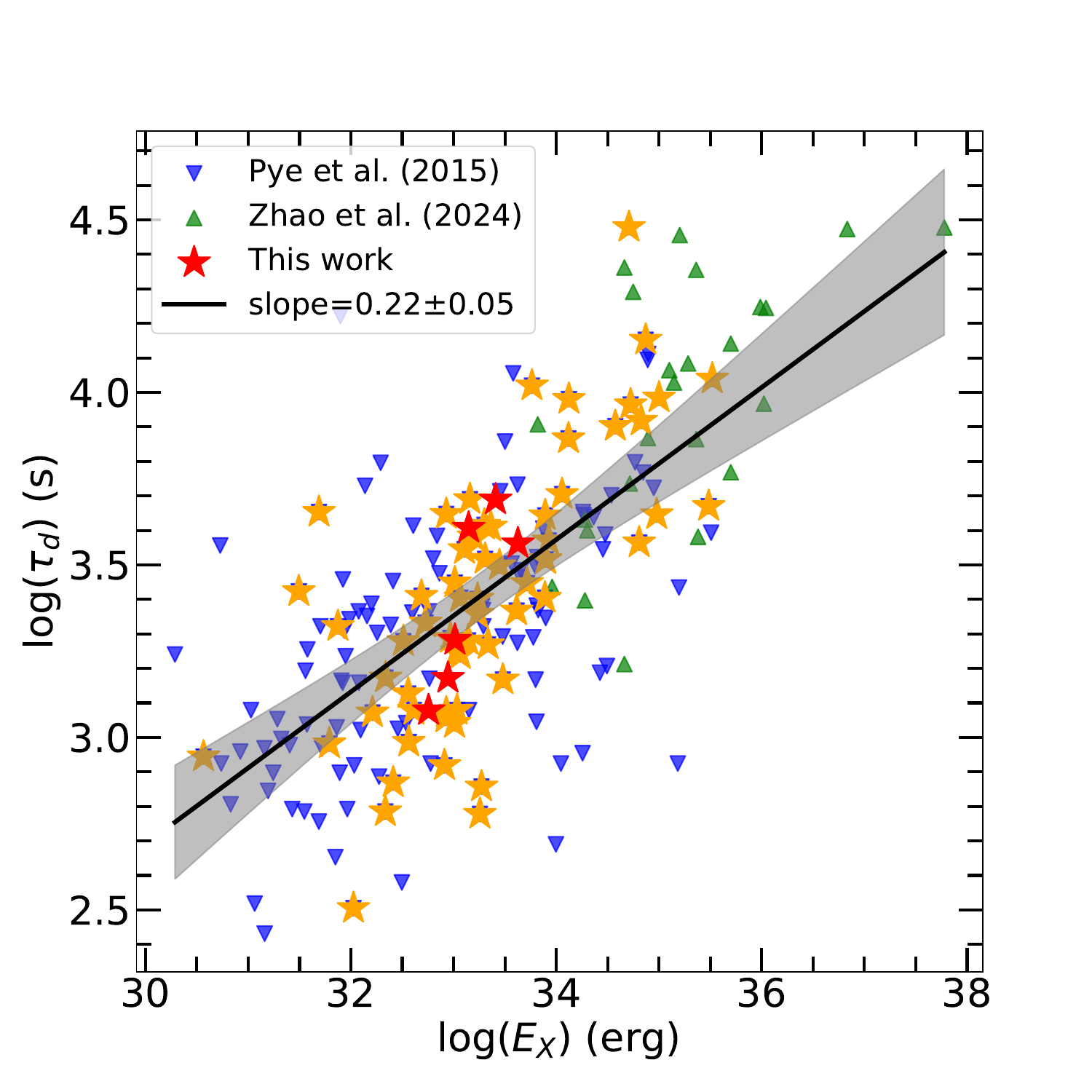}}
\caption{Flare energy and temporal flare parameters relations. The K-dwarf sub-sample from both studies \citep{2015A&A...581A..28P,2024ApJ...961..130Z} is denoted by orange star markers to clearly distinguish these targets from the broader stellar sample. (a) Flare energy vs flare peak luminosity relation, (b) Flare energy vs flare duration relation, (c) Flare energy vs flare rise times relation, and (d) Flare energy vs flare decay time.}
    \label{fig:relations}
\end{figure*}

\subsection{Flare parameters: comparison with Sun and other active stars}
We found that the flaring parameters derived in table \ref{tab:loopparam} lie within the range of those observed in other stellar flares from different active stars \citep[see][and references therein]{2003A&A...411..587V, 2007MNRAS.379.1075R, 2008MNRAS.387.1627P,2012MNRAS.419.1219P, 2015AJ....149...47P,2022A&A...666A.198P,2023MNRAS.518..900K,2025NewA..11402295S}.
Based on the study of super-hot solar flares of M and X class, \cite{2014ApJ...781...43C} has reported a statistical relationship between GOES class and the maximum flare temperature.  

To compare current studied flares with M and X-class super-hot solar flares, we converted the  X-ray flux in the 0.3-10.0 keV band to the GOES 1-8 $\AA$ flux using a multi-component model in the WebPIMMS \footnote{https://heasarc.gsfc.nasa.gov/cgi-in/Tools/w3pimms/w3pimms.pl} tool by adopting 3T-APEC parameter values of each flare segment. The flare energy is then calculated by integrating the GOES flux across all flare segments for all flares. The resulting flare energy in the GOES band for the flares F1-F6 is estimated to be in the range 4.9$\times$10$^{31}$ to 3.5$\times$10$^{32}$ erg. By scaling the peak GOES flux from the star to a distance of 1 AU, the studied flares lie in the X190-X432 class category of solar flares. 
We compared the GOES class of all stellar flares studied currently to the possible maximum temperature using the relation given by \cite{2014ApJ...781...43C}. All the studied flares agree with the predicted maximum temperature of the corresponding GOES class, except the flare F3, where the temperature is found to be hotter than predicted from the solar flare study. The high temperature might be due to the relatively longer rise time of the flare F3, which gives the plasma enough time to heat up to such high temperatures. The inferred coronal densities are in the range of 10$^{11}$ cm$^{-3}$, which are comparable to energetic X-class solar flares, but we note that the volume of flaring plasma is orders of magnitude higher than in the solar case. This is mainly due to the higher emission measure of stellar coronae, which exceeds that of the solar corona. The measured loop length and minimum magnetic field required to trap plasma within the loop during the flare peak are found to be higher than those of the X-class solar flares.

 To compare the flare properties of V834 Tau, BY Dra, and LQ Hya with flares from other active stars, we have taken the values of flare energy or total energy (E$_X$), peak X-ray luminosity (L$_p$), flare duration (T$_d$), $\tau_r$ and $\tau_d$ from the studies by \cite{2015A&A...581A..28P} and \cite{2024ApJ...961..130Z}. While \cite{2015A&A...581A..28P} sampled diverse spectral classes and evolutionary ages, the study by \cite{2024ApJ...961..130Z} was restricted to G-K type main-sequence stars.  The positive correlation observed in both studies thus appears to be a common property of flares, irrespective of whether they occur on stars of different spectral classes with varying magnetic activity. 
 The Figure \ref{fig:relations} shows the relation of E$_X$ with L$_p$, $\tau_r$ and $\tau_d$ with inclusion of flares from the current study.
The L$_p$ and E$_X$ are found to be strongly correlated with the Pearson correlation coefficient of 0.94, whereas E$_X$ is moderately correlated with T$_d$ and $\tau_r$, with correlation coefficients of 0.59 and 0.46, respectively.  Conversely, to the $\tau_r$ correlation, the $\tau_d$ correlation with E$_X$ is found to be stronger, with a correlation coefficient of 0.7. In all cases, the probability of no correlation is found to be less than 0.1~\%. Since the stars analyzed in this study are K-type, we highlight the K-dwarf subsample with orange star-shaped markers to clearly distinguish them within the full sample.
We then fitted a power-law model to the above relations by performing a linear fit in log-log space. The slope was calculated using standard $\chi^2$ minimization, and the bootstrap method was used to quantify the error in the slope. Briefly, in the bootstrap method,  a million random samples are generated from the original dataset, and the slope is calculated for each random sample. From the distribution of slope values, 3$\sigma$ was taken as the error in the fitted slope from the original data. In Figure \ref{fig:relations},  the best-fit slope for each relation is shown by a black continuous line,  whereas the 3$\sigma$ upper and lower bounds for the slope are plotted as the shaded gray region. The resulting relations are as follows:
\begin{eqnarray}
     L_{p}\propto E_{X}^{0.79\pm0.09}\\
    T_{d}\propto E_{X}^{0.17\pm0.05} \\ 
    \tau_r \propto E_{X}^{0.15\pm0.06} \\
    \tau_d \propto E_{X}^{0.22\pm0.05}
\end{eqnarray}
We find a similar relationship between different parameters to that found in the individual studies of \cite{2024ApJ...961..130Z} and \cite{2015A&A...581A..28P}, but combining them makes the relationship more precise. These relations demonstrate that $L_P$ does not scale linearly with $E_X$. This deviates from the expectations of the exponential rise and decay (ERD) model of flares, which assumes a linear correlation when T$_d$ remains constant\footnote{Total energy of the flare $E_X$  $\approx$ L$_p$$\times$(T$_d$), and  $T_d =\tau_r +  \tau_d$. For constant T$_d$, L$_p$ $\propto$ E$_X^{1}$}. Instead, the slope is $\sim$0.8 (less than unity) for log($L_{p}$)--log($E_{X}$) relation, considering the ERD model such a slope in $L_p$ and $E_X$ demands T$_d$ to be proportional to E$_X^{\sim 0.2}~ $\footnote{If L$_p$ $\propto$E$_X^\alpha$ and ($T_d$)$ \propto E_X^\beta$, then $\alpha+\beta=1$ to ensure E$_X$ $\approx$ L$_p$$\times$ T$_d$}. The moderate correlation of the T$_d$ - E$_X$ relationship suggests weak dependence of T$_d$ on E$_X$ and the exponent of $\sim$0.2 further shows the underlying morphology of flares remains governed by the ERD model. These relations show that the energy of a flare is mainly governed by its peak luminosity. Furthermore, defining the flare impulsiveness ($I$) as the ratio of peak luminosity to duration ($I = L_p / T_d$), we find that it scales with the X-ray energy as  \textit{I}=L$_p$/T$_d$ $\propto$ E$_X^{0.6\pm0.1}$. The positive relationship between the impulsiveness parameter and flare energy indicates that flares become more impulsive as flare energy increases.



We also examined the spectral class distribution of the flaring sample employed in these flare relationships. We found that K-type stars exhibit a spread comparable to that of the entire sample (see Figure \ref{fig:relations}). This observation suggests that stellar flares possess similar characteristics regardless of whether they occur on K-type stars or any other spectral class.

\section{Summary and Conclusions}\label{sec:conc}
We have carried out a detailed study of the X-ray quiescent and flaring states of three K-type active dwarfs, V834 Tau, BY Dra, and LQ Hya, which lie in the saturation region of the activity-rotation relationship. 
The quiescent X-ray spectra of all three stars are well described by two-temperature plasma models, indicating broadly similar coronal thermal structures. However, the emission measures of the cool and hot plasma components differ from star to star. Elemental abundance analysis reveals significant iron depletion in the coronae, by factors of 5–10 compared to the respective photospheric values, with the overall abundance pattern exhibiting a clear inverse-FIP effect. A total of six flares are analyzed in the current study, with GOES classifications ranging from X190 to X432. LQ Hya exhibits recurrent superflares near the same rotational phase, suggesting the presence of long-lived, complex magnetic structures at specific stellar longitudes. The occurrence of flares near the periastron passage of the BY Dra binary suggests that they may originate within the inter-binary magnetic field. Using time-resolved spectroscopy, we determine the physical properties of the flares and find that they reach extreme peak temperatures up to 100 MK. The loop length of the flaring plasma is estimated to be in the range of  $10^{10-11}$ cm, whereas the total energy released during the X-ray flares from these stars is of the order of $10^{33}$ erg.  A comparison of flare parameters with those of other stars and the Sun is made and found to be well within the range of statistical studies. 

\section*{Acknowledgements} 

The work presented here is based on the observations of the
XMM-Newton satellite, an ESA science mission with instruments and contributions directly funded by ESA Member States and NASA.

\bibliography{example}{}
\bibliographystyle{aasjournalv7}

\appendix

\section{Time resolved spectroscopy}
\renewcommand{\thetable}{A\arabic{table}}
\setcounter{table}{0}
The Table \ref{tab:TRS} shows best-fit spectral parameters for V834 Tau, BY Dra, and LQ Hya for different time segments of each observation. 
\startlongtable
\tabletypesize{\scriptsize}
\begin{deluxetable}{lllllllllll}
    \tablecaption{Best fit spectral parameters for stars.}\label{tab:TRS}
    \tablehead{
  \colhead{Seg} &    \colhead{Ts:Tp} &   \colhead{Z}&     \colhead{  T$_{1}$ }&        \colhead{T$_{2}$ }&   \colhead{ T$_{3}$}&        \colhead{ EM$_{1}$} &     \colhead{ EM$_{2}$} &  \colhead{EM$_{3}$}&    \colhead{Lx} &          \colhead{$\chi^2_\nu$ / d.o.f.} \\
  \colhead{(1)} &    \colhead{(2)} &   \colhead{(3)}&     \colhead{ (4) }&        \colhead{(5)}&   \colhead{(6)}&        \colhead{ (7)} &     \colhead{ (8)} &  \colhead{(9)}&    \colhead{(10)} &          \colhead{(11)} 
    }
    \startdata
     \multicolumn{11}{c}{V834 Tau}\\
     \hline
F1PE1&0.0:4.695&$0.19_{-0.01}^{+0.02}$&$0.26_{-0.01}^{+0.01}$&$0.93_{-0.02}^{+0.02}$&---&$1.3_{-0.1}^{+0.1}$&$1.0_{-0.1}^{+0.1}$&---&$1.71_{-0.02}^{+0.02}$&1.04/229\\ 
F1PE2&4.695:9.699&$0.20_{-0.01}^{+0.02}$&$0.27_{-0.01}^{+0.01}$&$0.94_{-0.02}^{+0.02}$&---&$1.1_{-0.1}^{+0.1}$&$0.9_{-0.1}^{+0.1}$&---&$1.52_{-0.01}^{+0.01}$&1.15/227\\ 
F1R1&9.801:10.192&$0.19_{-0.01}^{+0.01}$&0.27*&0.94*&$\ge$1.91&1.2*&0.97*&$0.1_{-0.1}^{+0.1}$&$1.69_{-0.05}^{+0.05}$&0.71/42\\ 
F1R2&10.192:10.9&$0.19_{-0.02}^{+0.01}$&0.27*&0.94*&$2.61_{-0.26}^{+0.30}$&1.2*&0.97*&$1.3_{-0.1}^{+0.1}$&$3.10_{-0.06}^{+0.06}$&0.80/113\\ 
F1P&10.9:11.506&$0.22_{-0.01}^{+0.02}$&0.27*&0.94*&$2.37_{-0.24}^{+0.31}$&1.2*&0.97*&$1.6_{-0.1}^{+0.1}$&$3.54_{-0.07}^{+0.07}$&1.07/115\\  F1D1&11.506:12.158&$0.28_{-0.02}^{+0.01}$&0.27*&0.94*&$1.68_{-0.13}^{+0.17}$&1.2*&0.97*&$1.0_{-0.1}^{+0.1}$&$2.99_{-0.06}^{+0.06}$&0.92/112\\ 
F1D2&12.158:12.999&$0.25_{-0.02}^{+0.01}$&0.27*&0.94*&$1.45_{-0.18}^{+0.30}$&1.2*&0.97*&$0.4_{-0.1}^{+0.1}$&$2.20_{-0.04}^{+0.04}$&1.02/107\\ 
F1PO1&13.1:15.388&$0.20_{-0.02}^{+0.03}$&$0.27_{-0.01}^{+0.01}$&$0.96_{-0.02}^{+0.03}$&---&$1.1_{-0.1}^{+0.1}$&$1.0_{-0.1}^{+0.1}$&---&$1.70_{-0.02}^{+0.02}$&1.15/189\\ 
\hline
\multicolumn{11}{c}{BY Dra}\\
\hline
F2P&0.0:0.443&$0.31_{-0.01}^{+0.01}$&0.27*&1.00*&$2.66_{-0.54}^{+0.93}$&3.2*&3.2*&$1.3_{-0.2}^{+0.2}$&$7.87_{-0.14}^{+0.14}$&1.00/124\\ 
F2D1&0.443:0.895&$0.32_{-0.02}^{+0.01}$&0.27*&1.00*&$2.64_{-0.50}^{+0.76}$&3.2*&3.2*&$1.1_{-0.2}^{+0.2}$&$7.60_{-0.13}^{+0.13}$&1.14/127\\  
F2D2&0.895:1.373&$0.30_{-0.02}^{+0.01}$&0.27*&1.00*&$1.78_{-0.40}^{+0.75}$&3.2*&3.2*&$0.9_{-0.2}^{+0.2}$&$6.98_{-0.12}^{+0.12}$&1.25/118\\ 
F2D3&1.373:1.917&$0.24_{-0.04}^{+0.04}$&$0.28_{-0.01}^{+0.01}$&$1.04_{-0.04}^{+0.03}$&---&$3.2_{-0.3}^{+0.3}$&$3.7_{-0.4}^{+0.4}$&---&$6.08_{-0.11}^{+0.11}$&1.04/124\\ 
F2D4&1.917:2.465&$0.24_{-0.04}^{+0.04}$&$0.26_{-0.01}^{+0.01}$&$1.04_{-0.04}^{+0.03}$&---&$3.6_{-0.3}^{+0.4}$&$3.5_{-0.4}^{+0.4}$&---&$6.00_{-0.11}^{+0.11}$&1.14/117\\ 
F2D5&2.465:3.0&$0.18_{-0.02}^{+0.03}$&$0.24_{-0.01}^{+0.02}$&$0.94_{-0.02}^{+0.03}$&---&$3.0_{-0.3}^{+0.3}$&$4.2_{-0.4}^{+0.4}$&---&$5.69_{-0.10}^{+0.10}$&0.99/113\\ 
F2PO1&3.0:5.912&$0.21_{-0.01}^{+0.01}$&$0.27_{-0.01}^{+0.01}$&$1.00_{-0.01}^{+0.01}$&---&$3.5_{-0.1}^{+0.1}$&$3.3_{-0.2}^{+0.2}$&---&$5.44_{-0.04}^{+0.04}$&1.28/259\\ 
F2PO2&5.912:8.827&$0.29_{-0.02}^{+0.02}$&$0.26_{-0.01}^{+0.01}$&$1.02_{-0.01}^{+0.02}$&---&$2.8_{-0.1}^{+0.1}$&$2.8_{-0.1}^{+0.1}$&---&$5.41_{-0.04}^{+0.04}$&1.18/252\\ 
F2PO3&8.827:11.399&$0.24_{-0.02}^{+0.02}$&$0.27_{-0.01}^{+0.01}$&$1.01_{-0.01}^{+0.01}$&---&$3.1_{-0.1}^{+0.2}$&$3.1_{-0.2}^{+0.2}$&---&$5.43_{-0.05}^{+0.05}$&1.08/245\\ 
F2PO4&11.4:12.277&$0.21_{-0.02}^{+0.03}$&$0.27_{-0.01}^{+0.01}$&$1.02_{-0.02}^{+0.03}$&---&$3.7_{-0.3}^{+0.3}$&$3.6_{-0.3}^{+0.3}$&---&$5.88_{-0.09}^{+0.09}$&1.00/169\\ 
F2PO5&12.277:13.202&$0.23_{-0.03}^{+0.03}$&$0.27_{-0.01}^{+0.01}$&$0.99_{-0.02}^{+0.02}$&---&$3.2_{-0.2}^{+0.3}$&$3.2_{-0.3}^{+0.3}$&---&$5.48_{-0.08}^{+0.08}$&1.09/176\\ 
F2PO6&13.202:14.209&$0.27_{-0.03}^{+0.04}$&$0.26_{-0.01}^{+0.01}$&$0.99_{-0.02}^{+0.02}$&---&$2.8_{-0.2}^{+0.2}$&$2.6_{-0.2}^{+0.2}$&---&$5.01_{-0.07}^{+0.07}$&1.13/171\\ 
F2PO7&14.209:15.182&$0.24_{-0.03}^{+0.03}$&$0.27_{-0.01}^{+0.01}$&$1.00_{-0.02}^{+0.02}$&---&$3.1_{-0.2}^{+0.2}$&$3.0_{-0.3}^{+0.3}$&---&$5.21_{-0.07}^{+0.07}$&1.08/174\\ 
F2PO8&15.182:16.196&$0.22_{-0.02}^{+0.03}$&$0.27_{-0.01}^{+0.01}$&$1.01_{-0.02}^{+0.03}$&---&$3.2_{-0.2}^{+0.3}$&$2.9_{-0.2}^{+0.3}$&---&$5.06_{-0.07}^{+0.07}$&1.00/175\\ 
F2PO9&16.196:17.146&$0.23_{-0.03}^{+0.03}$&$0.27_{-0.01}^{+0.01}$&$0.99_{-0.02}^{+0.02}$&---&$3.2_{-0.2}^{+0.3}$&$3.2_{-0.3}^{+0.3}$&---&$5.37_{-0.08}^{+0.08}$&1.03/177\\ 
F2PO10&17.146:18.04&$0.20_{-0.02}^{+0.03}$&$0.27_{-0.01}^{+0.01}$&$1.00_{-0.02}^{+0.02}$&---&$3.1_{-0.2}^{+0.2}$&$3.9_{-0.3}^{+0.3}$&---&$5.75_{-0.08}^{+0.08}$&0.97/174\\ 
F2PO11&18.04:18.982&$0.24_{-0.02}^{+0.04}$&$0.27_{-0.01}^{+0.01}$&$0.97_{-0.03}^{+0.02}$&---&$2.9_{-0.2}^{+0.2}$&$3.1_{-0.3}^{+0.3}$&---&$5.38_{-0.08}^{+0.08}$&0.90/173\\ 
F2PO12&18.982:19.935&$0.26_{-0.03}^{+0.04}$&$0.27_{-0.01}^{+0.01}$&$1.03_{-0.03}^{+0.02}$&---&$2.9_{-0.2}^{+0.2}$&$3.0_{-0.3}^{+0.3}$&---&$5.35_{-0.08}^{+0.08}$&1.01/178\\ 
F2PO13&19.935:20.7&$0.18_{-0.03}^{+0.02}$&$0.25_{-0.01}^{+0.01}$&$0.98_{-0.03}^{+0.03}$&---&$3.5_{-0.3}^{+0.3}$&$3.8_{-0.3}^{+0.3}$&---&$5.37_{-0.09}^{+0.09}$&1.10/147\\ 
\hline
\multicolumn{11}{c}{LQ Hya (Epoch-I)}\\
\hline
F3PE1&0.0:1.056&$0.17_{-0.02}^{+0.02}$&$0.28_{-0.01}^{+0.01}$&$1.00_{-0.02}^{+0.02}$&---&$3.3_{-0.2}^{+0.2}$&$3.9_{-0.3}^{+0.3}$&---&$5.28_{-0.07}^{+0.07}$&0.99/191\\ 
F3PE2&1.056:2.327&$0.15_{-0.02}^{+0.01}$&$0.28_{-0.01}^{+0.01}$&$0.98_{-0.02}^{+0.02}$&---&$3.3_{-0.2}^{+0.2}$&$4.5_{-0.3}^{+0.3}$&---&$5.52_{-0.06}^{+0.06}$&0.99/204\\ 
F3PE3&2.327:3.621&$0.16_{-0.01}^{+0.02}$&$0.26_{-0.01}^{+0.01}$&$0.96_{-0.02}^{+0.02}$&---&$3.2_{-0.2}^{+0.2}$&$4.2_{-0.3}^{+0.3}$&---&$5.38_{-0.06}^{+0.06}$&1.09/203\\ 
F3PE4&3.621:4.987&$0.14_{-0.02}^{+0.01}$&$0.28_{-0.01}^{+0.01}$&$0.94_{-0.02}^{+0.02}$&---&$3.5_{-0.2}^{+0.2}$&$4.1_{-0.3}^{+0.3}$&---&$5.10_{-0.06}^{+0.06}$&1.11/202\\ 
F3PE5&5.1:6.497&$0.16_{-0.02}^{+0.02}$&$0.30_{-0.01}^{+0.01}$&$1.00_{-0.02}^{+0.02}$&---&$3.2_{-0.2}^{+0.2}$&$3.9_{-0.2}^{+0.2}$&---&$5.19_{-0.06}^{+0.06}$&1.15/205\\ 
F3PE6&6.497:7.878&$0.15_{-0.02}^{+0.01}$&$0.29_{-0.01}^{+0.01}$&$0.98_{-0.02}^{+0.02}$&---&$3.7_{-0.2}^{+0.2}$&$4.4_{-0.3}^{+0.3}$&---&$5.66_{-0.06}^{+0.06}$&1.12/212\\ 
F3PE7&7.878:9.367&$0.17_{-0.02}^{+0.02}$&$0.28_{-0.01}^{+0.01}$&$0.97_{-0.02}^{+0.02}$&---&$3.3_{-0.2}^{+0.2}$&$3.8_{-0.2}^{+0.3}$&---&$5.24_{-0.06}^{+0.06}$&1.03/208\\
F3PE8&9.367:10.887&$0.18_{-0.02}^{+0.02}$&$0.27_{-0.01}^{+0.01}$&$0.97_{-0.03}^{+0.02}$&---&$3.2_{-
0.2}^{+0.2}$&$3.6_{-0.2}^{+0.3}$&---&$5.12_{-0.06}^{+0.06}$&1.12/208\\ 
F3PE9&10.887:16.349&$0.14_{-0.00}^{+0.01}$&$0.28_{-0.01}^{+0.01}$&$0.95_{-0.01}^{+0.01}$&---&$3.4_{-0.1}^{+0.1}$&$4.3_{-0.1}^{+0.1}$&---&$5.31_{-0.03}^{+0.03}$&1.28/339\\ 
F3PE10&16.349:19.692&$0.14_{-0.01}^{+0.01}$&$0.31_{-0.01}^{+0.01}$&$0.99_{-0.01}^{+0.01}$&---&$3.5_{-0.1}^{+0.1}$&$4.4_{-0.2}^{+0.2}$&---&$5.52_{-0.04}^{+0.04}$&1.17/278\\ 
F3R1&19.692:21.813&$<0.16$&0.28*&0.97*&$4.24_{-0.65}^{+0.96}$&3.4*&4.1*&$0.8_{-0.1}^{+0.1}$&$6.78_{-0.06}^{+0.06}$&1.19/265\\ 
F3P&21.813:23.576&0.18*&0.28*&0.97*&$2.36_{-0.19}^{+0.22}$&3.4*&4.1*&$1.8_{-0.1}^{+0.1}$&$8.06_{-0.07}^{+0.07}$&1.11/272\\  
F3D1&23.576:25.611&$>0.17$&0.28*&0.97*&$2.79_{-0.42}^{+0.63}$&3.4*&4.1*&$0.5_{-0.1}^{+0.1}$&$6.73_{-0.06}^{+0.06}$&1.24/255\\ 
 F3D2&25.611:27.866&$0.16_{-0.02}^{+0.01}$&$0.28_{-0.01}^{+0.01}$&$0.97_{-0.02}^{+0.02}$&---&$4.0_{-0.2}^{+0.2}$&$4.3_{-0.2}^{+0.2}$&---&$5.90_{-0.05}^{+0.05}$&1.09/242\\ 
 F4R1&28.0:28.422&$0.15_{-0.03}^{+0.02}$&$0.31_{-0.02}^{+0.03}$&$0.98_{-0.04}^{+0.04}$&---&$3.7_{-0.3}^{+0.4}$&$4.3_{-0.4}^{+0.5}$&---&$5.62_{-0.11}^{+0.11}$&1.19/97\\ 
F4P&28.422:28.821&$0.17_{-0.01}^{+0.01}$&0.28*&0.97*&$3.40_{-0.94}^{+2.55}$&3.4*&4.1*&$1.2_{-0.3}^{+0.3}$&$7.53_{-0.14}^{+0.14}$&1.16/115\\ 
F4D1&28.821:29.256&$0.17_{-0.01}^{+0.01}$&0.28*&0.97*&$\ge4.07$&3.4*&4.1*&$0.4_{-0.1}^{+0.1}$&$6.61_{-0.12}^{+0.12}$&0.95/112\\ 
 F4D2&29.256:29.739&$0.16_{-0.03}^{+0.02}$&$0.30_{-0.02}^{+0.02}$&$1.00_{-0.03}^{+0.04}$&---&$3.7_{-0.3}^{+0.3}$&$4.2_{-0.4}^{+0.4}$&---&$5.68_{-0.10}^{+0.10}$&1.02/104\\ 
 F4D3&29.739:29.999&$0.14_{-0.03}^{+0.04}$&$0.35_{-0.04}^{+0.06}$&$1.00_{-0.05}^{+0.06}$&---&$3.4_{-0.4}^{+0.4}$&$4.4_{-0.6}^{+0.6}$&---&$5.61_{-0.14}^{+0.14}$&1.06/66\\ 
 \multicolumn{11}{c}{LQ Hya (Epoch-II)}\\
F5P&0.0:0.447&$0.18_{-0.01}^{+0.01}$&0.30*&1.01*&$1.76_{-0.22}^{+0.32}$&3.6*&4.6*&$1.7_{-0.3}^{+0.3}$&$8.03_{-0.13}^{+0.13}$&0.89/135\\  
F5D1&0.447:0.928&$0.17_{-0.01}^{+0.01}$&0.30*&1.01*&$2.41_{-0.52}^{+0.92}$&3.6*&4.6*&$1.0_{-0.2}^{+0.3}$&$7.47_{-0.12}^{+0.12}$&1.08/139\\ 
F5D2&0.928:1.458&$0.13_{-0.01}^{+0.02}$&0.30*&1.01*&$0.70_{-0.09}^{+0.07}$&3.6*&4.6*&$1.1_{-0.5}^{+0.5}$&$6.52_{-0.11}^{+0.11}$&0.98/134\\  F5D3&1.458:2.003&$0.16_{-0.01}^{+0.01}$&0.30*&1.01*&$0.24_{-0.06}^{+0.08}$&3.6*&4.6*&$0.6_{-0.3}^{+0.3}$&$6.34_{-0.10}^{+0.10}$&0.99/140\\ 
 F1D4&2.003:2.573&$0.14_{-0.02}^{+0.03}$&$0.31_{-0.02}^{+0.03}$&$0.96_{-0.03}^{+0.04}$&---&$3.9_{-0.3}^{+0.3}$&$4.7_{-0.5}^{+0.5}$&---&$5.99_{-0.10}^{+0.10}$&1.01/136\\ 
 F5D5&2.573:3.149&$0.09_{-0.01}^{+0.02}$&$0.42_{-0.06}^{+0.08}$&$0.95_{-0.04}^{+0.05}$&---&$3.8_{-0.5}^{+0.5}$&$5.5_{-0.6}^{+0.6}$&---&$5.99_{-0.10}^{+0.10}$&0.98/139\\ 
 F5D6&3.149:3.7&$0.14_{-0.02}^{+0.02}$&$0.31_{-0.02}^{+0.03}$&$1.00_{-0.03}^{+0.04}$&---&$3.6_{-0.3}^{+0.3}$&$4.8_{-0.5}^{+0.4}$&---&$5.96_{-0.10}^{+0.10}$&0.85/129\\ 
 F5PO1&3.8:5.56&$0.13_{-0.01}^{+0.02}$&$0.33_{-0.01}^{+0.01}$&$1.04_{-0.02}^{+0.02}$&---&$4.0_{-0.2}^{+0.2}$&$4.9_{-0.2}^{+0.2}$&---&$6.13_{-0.06}^{+0.06}$&1.26/236\\ 
 F5PO2&5.56:7.431&$0.14_{-0.01}^{+0.01}$&$0.28_{-0.01}^{+0.01}$&$0.98_{-0.02}^{+0.02}$&---&$3.5_{-0.2}^{+0.2}$&$4.8_{-0.2}^{+0.2}$&---&$5.76_{-0.05}^{+0.05}$&1.15/237\\ 
 F5PO3&7.431:9.353&$0.15_{-0.01}^{+0.02}$&$0.30_{-0.01}^{+0.01}$&$1.01_{-0.02}^{+0.02}$&---&$3.5_{-0.2}^{+0.2}$&$4.3_{-0.2}^{+0.2}$&---&$5.59_{-0.05}^{+0.05}$&1.15/232\\ 
 F5PO4&9.353:11.2&$0.15_{-0.01}^{+0.02}$&$0.30_{-0.01}^{+0.01}$&$1.01_{-0.02}^{+0.02}$&---&$3.6_{-0.2}^{+0.2}$&$4.4_{-0.2}^{+0.2}$&---&$5.74_{-0.05}^{+0.05}$&1.18/235\\ 
 F6R1&11.501:12.004&$0.11_{-0.01}^{+0.02}$&$0.27_{-0.03}^{+0.03}$&$0.96_{-0.03}^{+0.04}$&---&$3.2_{-0.4}^{+0.4}$&$6.0_{-0.6}^{+0.6}$&---&$6.06_{-0.11}^{+0.11}$&1.01/118\\ 
F6P&12.004:12.428&$0.15_{-0.01}^{+0.01}$&0.30*&1.01*&$2.40_{-0.30}^{+0.36}$&3.6*&4.6*&$3.0_{-0.3}^{+0.3}$&$9.05_{-0.15}^{+0.15}$&1.13/143\\ 
F6D1&12.428:13.414&$<0.18$&0.30*&1.01*&$4.63_{-1.09}^{+2.21}$&3.6*&4.6*&$0.9_{-0.1}^{+0.1}$&$7.73_{-0.09}^{+0.09}$&1.01/216\\ 
F6D2&13.414:14.425&$0.16_{-0.01}^{+0.01}$&0.30*&1.01*&$2.42_{-0.37}^{+0.51}$&3.6*&4.6*&$1.0_{-0.2}^{+0.2}$&$7.23_{-0.08}^{+0.08}$&1.13/218\\ 
F6D3&14.425:14.938&$0.15_{-0.01}^{+0.01}$&0.30*&1.01*&$2.07_{-0.48}^{+0.95}$&3.6*&4.6*&$1.1_{-0.3}^{+0.3}$&$7.08_{-0.12}^{+0.12}$&1.01/134\\ 
F6D4&14.938:15.4&$<0.16$&0.30*&1.01*&$2.67_{-0.73}^{+1.51}$&3.6*&4.6*&$0.8_{-0.2}^{+0.2}$&$6.96_{-0.12}^{+0.12}$&1.14/124\\ 
\enddata
    \tablecomments{
     (1): Name of the segment, (2): Ts and Tp are start and stop times of particular segment from the start of observation, (3): Z is abunundances in units of Z$_\odot$, (4)-(6):    the temperatures ($T_1$, $T_2$, $T_3$) are in units of keV, (7)-(9): emission measures (EM$_1$, EM$_2$, EM$_3$) are in units of 10$^{52}$cm$^{-3}$,
     (10): the luminosity L$_X$ in energy band of  0.3-10.0 keV is in units of  $10^{29}$ ergs$^{-1}$, and (12): reduced $\chi^2$ and degree of freedom
     }
\end{deluxetable}

\end{document}